\documentclass[useAMS,usenatbib]{mn2e}

\usepackage{graphicx}
\usepackage{amssymb}
\usepackage{pstricks}
\usepackage{longtable,lscape}

\def \be {\begin{equation}}
\def \ee {\end{equation}}

\def \be {\begin{equation}}

\def \ee {\end{equation}}

\title{{ Migration of Earth-size}  planets  in 3D radiative discs}
\author[ E. Lega, A. Crida, B. Bitsch, A. Morbidelli  ] { 
E. Lega$^{1}$, \thanks{E-mail: elena@oca.eu (EL); crida@oca.eu (AC);   bbitsch@oca.eu (BB); morby@oca.eu (AM)}, 
A. Crida ${^1}$,   B. Bitsch $^{1}$, A.Morbidelli ${^1}$ \\
$^{1}$ Universit\'e Nice Sophia Antipolis, CNRS UMR 7293,  Observatoire de la
C\^ote d'Azur,   Bv. de l'Observatoire, \\CS~34229,  06304 Nice cedex 4, 
France.\\
}
\begin{document}
\maketitle
\date{Accepted .... Received ....; in original form ....}

\pagerange{\pageref{firstpage}--\pageref{lastpage}} \pubyear{2013}

\maketitle
\label{firstpage}

\begin{abstract}
{ In this paper, we address the migration of small mass planets in 3D radiative disks. Indeed, migration of small planets is known to be too fast inwards in locally isothermal conditions. However, thermal effects could reverse its direction, potentially saving planets in the inner, optically thick parts of the protoplanetary disc.  This effect has been seen for masses larger than 5 Earth masses, but the minimum mass for this to happen has never been probed numerically, although it is of crucial importance for planet formation scenarios.
We have extended the hydro-dynamical code FARGO  to 3D, with thermal diffusion. With this code, we perform simulations of embedded planets   down to 2 Earth masses. For a set of discs parameters for which outward migration has been shown in the range of $[5, 35]$ Earth masses, we find that the transition to inward migration occurs for masses in the range $[3, 5]$ Earth masses. The transition appears to be due to an unexpected phenomenon:  the formation of an asymmetric cold and dense finger of gas driven by circulation and libration streamlines. We recover this phenomenon in 2D simulations where we control the cooling effects of the gas through a simple modeling of the energy equation.} 
 \end{abstract}

\begin{keywords}
Protoplanetary discs, hydrodynamics, planetary systems, methods: numerical.
\end{keywords}

\section{Introduction}
The cores of giant planets and possibly many super-earths  
 form and migrate whitin a protoplanetary disc.
In a locally isothermal disc, planets with masses $m_p \la 10-15$ Earth masses
($M_{\earth}$)
are expected to undergo type-I migration towards the central star on a timescale proportional to $1/m_p$  (\citet{Tanaka02}) , which can be shorter than the disc lifetime for
$m_p \ga 1 M_{\earth} $. If this were true, planetary cores should not survive in the disc,  in evident contrast with the observation
that many planetary systems, including our own, have planets that are not even very close to the star.
\par
Many efforts have been done to solve this puzzling problem.
 Different mechanisms have been suggested  to halt inward migration such as magnetic field effects in turbulent discs (\citet{Baruteau11,Guilet13} and references therein) or density trap (\citet{Masset02,MassetMorby06,MorbyCrida08}). But, perhaps the most promising mechanism appears when
  considering the effect of energy transfer within the disc.  Studies
have  shown that non-isothermal effects can change the direction 
of the type-I migration (\citet{PaardeMelle06,BarMass08,KleyCri08,KBK09,MasCas09,MassetCasoli10,Paaretal10}). \par
Precisely,  the migration in the inner 
part of a  radiative disc can be directed outward, 
while it remains directed  inward in the outer disc. This establishes the existence of a critical radius where
migration vanishes, towards which  planetary cores  migrate from both the inner and the outer part of the disc. Therefore, the zero  migration location
 acts as  a planet trap (\citet{Lyraetal10}). \par
The   torque exerted by the protoplanetary disc has two main contributions:  i) the so-called Lindblad torque due to the spiral arms launched by the planet  in the disc which is not affected by the equation of state \footnote{Actually, it scales with $\gamma \Gamma_0$, with $\Gamma_0$ given in \ref{gamma0}.},  and ii) the corotation torque caused by material librating in the horseshoe region.  
When considering radiative effects it has been shown that the corotation
torque contribution can be positive and possibly dominate over the negative
Lindblad torque.
 In a recent paper, \citet{Paaretal11}, have provided a detailed formula
for the torque which was calibrated with a 2D hydrodynamical model for low mass planets ($5 M_{\earth}$).
However, the corotation effect depends on  the disc parameters and on the
mass of the planet.  Decreasing the planetary mass, the horseshoe {U-turn} timescale {increases and, if it} becomes longer than the radiative diffusion timescale{, the} disc behaves similarly to the isothermal case and the positive torque contribution fades.
\par The  aim of this paper is to  investigate  the total torque acting on low mass planets kept on fixed orbits 
and possibly  find the critical mass for which the positive contribution
of the corotation torque becomes smaller than the Lindblad torque. To this purpose we have extended  the FARGO code of \citet{Masset00}, \citet{Masset00b} to 3 dimensions  introducing an energy equation to provide a realistic modeling of radiative effects as in \citet{KBK09}.
We have validated the 3 dimensional version of FARGO (FARGO with Colatitude Added, FARGOCA hereafter) by 
reproducing some results published in the literature (see Appendix) . \par
We have computed the total torque on planets with mass ranges in the interval 
$[2,10]$ Earth masses for a specific set of disc parameters
 and we found that a transition from positive to negative total torque occurs for a planet with mass between 3 and 5  Earth masses. We remark that for the same disc 
parameters \citet{KBK09} found positive total torque for planetary masses in the
range $[5,35]$ Earth masses.  When  applying the torque formula 
provided in  \citet{Paaretal11} to our disc parameters we  find a transition from outward to inward migration occurring between  2.7 and 3  Earth masses. \par
Despite the good agreement between model and simulations it is important to
study the mechanism that causes the transition  since in the case
 of \citet{Paaretal11} the results were calibrated with 2D simulations
while our work is done with numerical simulations of a full radiative 3D model. For this purpose we have analyzed the  torque suffered by the planet from every radial ring of the disc.   We have found a  new and unexpected feature:  a negative contribution to the torque that is not seen for planetary masses larger than $m_p \ga 5M_{\earth}$. 
This is due to an overdensity of gas behind the planet location just outside of its orbit.

  We have recovered the same effect in 2D simulations where the energy equation has a  simple cooling prescription in which the cooling time appears as a parameter.
Despite its simplicity, the 2D model allows us  to confirm the result and 
to study its dependence on the gas cooling timescale. 
\par

 The paper is organized as follows: the physical modelling is presented in Section 2, in Sect. 3   total torque calculations  on small mass planets are described.
We provide in Sect. 4 a detailed analysis of the local torque
acting on the planets as well as a mechanism for
 the origin of the negative and positive
 local contributions to the torque  that we observe on small mass planets.
We validate the proposed mechanism in Sect. 5 using  2D simulations.
In Sect. 6 we explain the origin of the asymmetry between negative and
positive torque contributions. 
 Conclusions are provided in Section 7.   In the appendix
 we provide test calculations  on the FARGOCA code.

\section{Physical modelling}
The protoplanetary disc is treated as a three dimensional non self-gravitating gas whose motion is described by the Navier-Stokes equations.
We use spherical coordinates $(r,\theta, \varphi)$ 
where $r$ is the radial distance from the the origin, $\theta$ the 
polar angle measured from the $z$-axis and $\varphi$ is the azimuthal coordinate
starting from the $x$-axis. The midplane of the disc is at the equator
$\theta = {\pi \over 2}$ and the origin of the coordinates is centered on the star. 
We work in a  coordinate system which rotates with angular velocity:
$$\Omega_p = \sqrt {G(M+m_p) \over {a_p}^3} \simeq \sqrt {GM \over {a_p}^3} $$
where $M$ is the mass of the central star 
and $a_p$ is the semi-major axis of a planet of mass $m_p$.
The gravitational influence of the planet on the disc 
 is modelled as in \cite{KBK09}
using a cubic-potential of the form:
\begin{equation}
\Phi _p = \left\lbrace \begin{array}{ll}
-{m_pG\over d} &  d > r_{\rm sm} \\
-{m_pG\over d}f({d\over r_{\rm sm}}) & d\leq r_{\rm sm}   
\end{array} \right.
\label{cubic}
\end{equation}
with $f({d\over r_{\rm sm}}) = \left [ \left( {d\over r_{\rm sm}}\right)^4-2\left( {d\over r_{\rm sm}}\right)^3+2{d\over r_{\rm sm}} \right]$;\\
  $d$ is the distance from the disc element to the planet, and $r_{\rm sm}$ the smoothing length:
$r_{\rm sm} = \alpha _{sm} R_H$, where $\alpha _{sm}$ is a smoothing parameter and
 $R_H$ is the Hill radius of the planet: 
$$R_H= a_p\sqrt[3]{m_p \over {3M}}\ .$$
Unless specified we take $\alpha _{sm}=0.5$ in our simulations.
The potential of Eq.\ref{cubic} is well suited to 3D simulations (\citet{KBK09}) and
different from the typical  $\epsilon$-potential used in 2 dimensional
models:
  \begin{equation}
\Phi _p^{\epsilon} = -{m_pG\over {\sqrt {d^2+\epsilon^2}}}
\label{smooth}
\end{equation}
Actually, in 2 dimensional models, Eq.\ref{smooth} allows to mimic the average influence  that the planet
would have  on the vertical gas column.
A value often used for the smoothing is $\epsilon = 0.7H$ where $H$ is the disc scaleheight{, but see \citet{Muller12} for a more detailed analysis.}
 \par
To the usual Navier-Stokes equations (see Appendix) we add
the equations for  the internal energy  modeled as in
 \cite{KBK09}:
\begin{equation}
\label{energy3D}
\left\lbrace \begin{array}{llr}
{\partial E \over \partial t}+ \nabla \cdot (E\vec v) & = & -p\nabla \cdot \vec v + Q^{+}  \\ &  & -\nabla \cdot D\nabla T \\
D = -{{\lambda c 4a_rT^3} \over {\rho \kappa}} 
\end{array} \right.
\end{equation}

where $E$ is the internal energy $E=\rho c_v T$, $T$ is the temperature of the disc and
$c_v$ is the specific heat at constant volume. On the right hand side
the first term denotes the compressional heating, $Q^{+}$ the expression for the viscous heating
and $ D\nabla T$ is the radiative flux.

The diffusion coefficient $D$ is provided by the flux-limited diffusion approach
(\citet{LevePom81}) with flux limiter   $\lambda$ 
 \footnote{ The expression that we use in the code is: 
	$$\lambda = 2/(3+\sqrt{9 +10s^2})$$ for $s\leq 2$ and
	$$\lambda = 10/(10s+9+\sqrt{81+180s}) $$ for $s>2$ with  $s= {4\over {\rho k}}{|\nabla T| \over T}$}
(\citet{Kley89}),
$c$ is the speed of light, $\kappa$ the Rosseland mean opacity and $a_r$ the radiation constant. We use the opacity law of \citet{BL94}.
The viscous heating can be found in  \cite{Mihalas}.\par

The system of 
equations is closed using an ideal gas equation of state: $P=R_{\rm gas}\rho T / \mu$
with  mean molecular weight $\mu=2.3 (g/mol) $ for standard solar mixture. 
Taking into account that:  $$E=R_{\rm gas}\rho T / \mu (\gamma-1)$$ this relates to the pressure as:
$$P = E(\gamma-1)$$ In the following we will use $\gamma =1.43${, where $\gamma$ is the adiabatic index}.
The sound  speed differs from the isothermal sound speed by a factor
$\sqrt \gamma$, that  is: $c_s = \sqrt{\gamma P/\rho}$. \par

We use units such that $G=1$ and ${M=}M_{\odot}=1$. The orbital period of a planet 
with semi-major axis $a_p=1$ is therefore $\tau =2\pi$. 
In the following we normalize the time $t$ by $\tau$ ,  so that $t$ corresponds to the number of orbits of the planet.  \par


\section{Transition from outward to inward migration: total torque}
The aim of this section is to follow the time evolution of the 
total torque until a stationary situation is reached. 
The stationary torque is computed for a range of 
 planetary masses with the goal of detecting for which mass   there is a
 transition from outward to inward migration (i.e. from positive to negative torque).
 We then compare the transition mass to that predicted in   
the formula provided by \citet{Paaretal11}.
We consider the disc setting of \cite{KBK09}, namely a disc of mass 
$M= 0.01 M_{\odot}$ with surface density $\Sigma(r)=\Sigma _0 (r/a_J)^{-1/2}$ 
with  $\Sigma_0= 6.76\times10^{-4}$ in code units ($2222\, \mbox{kg\, m}^{-2}$),
initial aspect ratio $h={ H\over r} = 0.05$ and extending from
$r_{min}\leq r \leq r_{max}$ with $r_{min}=0.4$ and $r_{max}=2.5$ in units of $a_J=5.2AU$.
In the vertical direction the disc extends from the midplane ($\theta \simeq 90^{\circ} $ to $7^{\circ}$ above the midplane, i.e. $\theta \simeq 83^{\circ} $). We do not study inclined planets orbits therefore we don't need to extend the disc below the midplane.   
{ We use the boundary conditions explained in the Appendix }. \par
Before placing planets in the 3D disc, we bring the latter to  radiative equilibrium. Because the disc is  
axisymmetric  in absence of the planet, we model
the disc in  2D, with  coordinates   $(r,\theta)$.
With our set of parameters, the opacity and a constant viscosity coefficient of
$10^{-5}$ in code units ({or $10^{11}\rm{m^2/s}$=}$10^{15} \rm{cm^2/s}$), the equilibrium
between viscous heating and radiative cooling reduces the aspect ratio of the disc from the initial value of $0.05$ to $0.037$ at $r=1$.  
Once the 2D equilibrium is achieved all the gas fields are expanded to 3D.
Planets of different masses are then embedded in such an equilibrium disc and
held on fixed  circular  orbits at $r=1$, $\varphi = \pi$, $\theta=\pi/2$ (midplane). Since the frame is corotating with the planet, the planet stays fixed at the corner of 8 cells.
We compute the  torque acting on  planets integrating over the whole disc until
the torque becomes almost constant (stationary state).
 \par
It is common to exclude the inner part of the Hill sphere of the planet from
the calculation of the gravitational torque acting on the planet.
This procedure allows to exclude the part of the disc that is  accreted by planets that form a circum-planetary disc.  However,
for small mass planets that don't have a circumplanetary disc 
this prescription is not  justified. In addition, in our computations
 the cells are symmetrically placed around the planet therefore the torque is
not affected by any artificial asymmetry that is sometimes avoided by using an Hill's cut. 
\par  In this 
section we discuss the total torque  without Hill cut. {  Unless specified, the total torque $\Gamma_{tot}$ is normalized with respect to $\Gamma_0$ (\citet{LinPap79}):
\begin{equation}
 \Gamma_0 = (q/h)^2\Sigma_p r^4_p\Omega^2_p 
\label{gamma0}
\end{equation}
where $\Sigma_p$ is the disc's surface density at the
planet location $r_p$ and $q$ is the mass ratio $q=m_p/M_{*}$.} \par

\begin{table*}
\begin{center}
\begin{minipage}{100mm}
\begin{tabular}{|lllll|}
\hline\noalign{\smallskip}
mass ($M_{\earth} $) & ($N_r$,$N_\theta,N_\varphi$) & $\alpha_{sm}$ & $n$ cells in $x_{hs}$  & \\
\noalign{\smallskip}\hline\noalign{\smallskip}
\hline\noalign{\smallskip}
  $10$ & $(254,42,754)$ & 0.5 & 4 & \\
  $5$ & $(262,32,768)$ & 0.5 & 3 & \\
  $5$ & $(359,50,1068)$ & 0.5 & 4 & \\
  $3$ & $(350,42,1025)$ & 0.5 & 3 & \\
  $3$ & $(464,60,1382)$ & 0.7 & 4 & \\
 $3$ & $(464,60,1382)$ & 0.5 & 4 & \\
 $3$ & $(464,60,1382)$ & 0.3 & 4 & \\
$3$ & $(464,60,1382)$ & 0.1 & 4 & \\
  $3$ & $(580,90,1727)$ & 0.5 & 5 & \\
 $2$  &  $(568,90,1696)$ & 0.5 & 4 & \\
\noalign{\smallskip}\hline
\end{tabular}
\caption{Simulations parameters.}
\label{table:tab1} 
\end{minipage}
\end{center}

\end{table*}

The resolution of our computational grid is chosen in order to have  in the radial direction $n$ grid-cells in the horseshoe region.
The half-width of planet's horseshoe region is given, in the isothermal disc approximation 
(\citet{MassetDangelo06}), by:
\begin{equation}
x_{hs}=1.16a_p\sqrt{q \over h} \ .
\label{xhs}
\end{equation}

In the following we consider different masses and different resolutions
$(N_r,N_\theta,N_\varphi)$
as shown in Table \ref{table:tab1}.

\begin{figure}
\includegraphics[height=6truecm,width=7truecm]{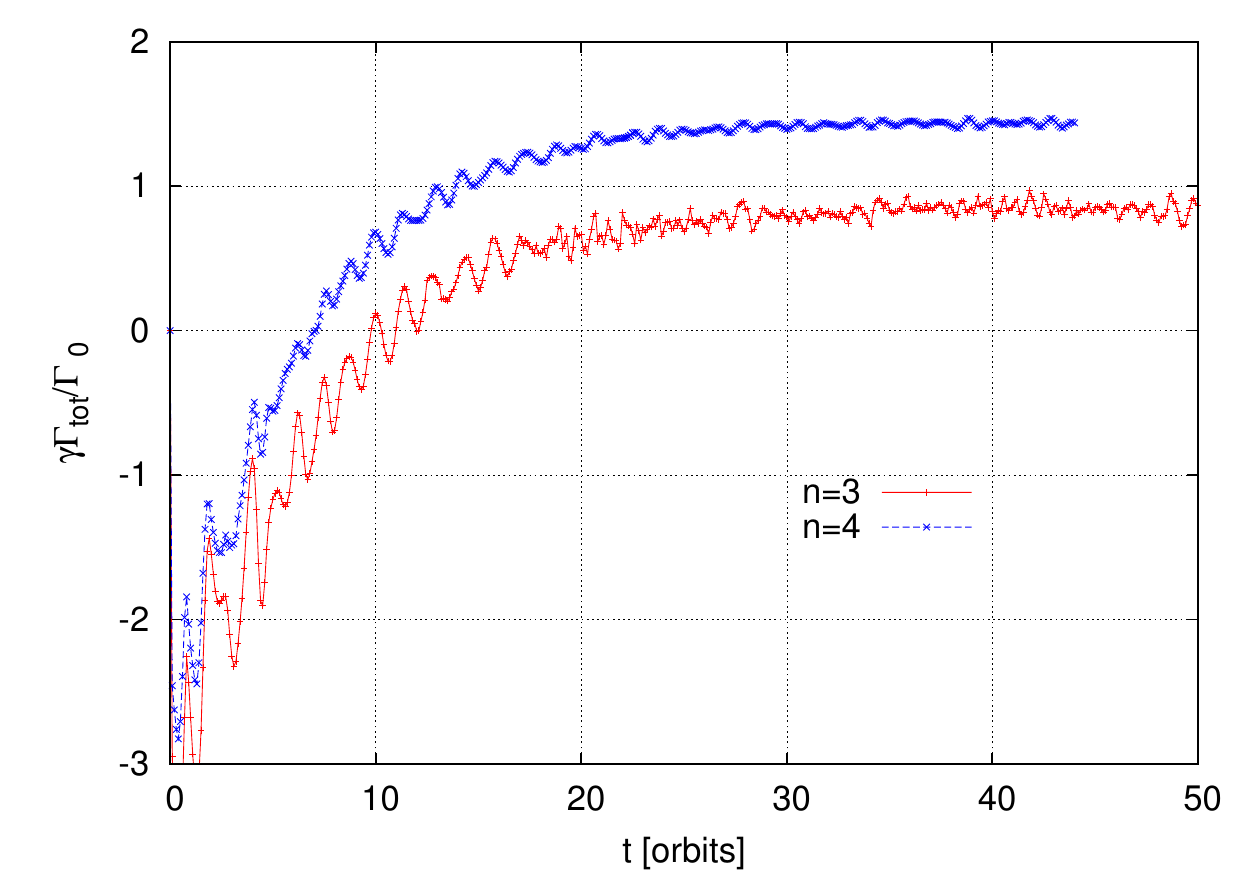}
\caption{Evolution of the total torque with time for a planet of $5M_{\earth}$ and two different resolutions corresponding to $n=3$ and $n=4$ cells in the horseshoe region. }
\label{m5tt}
\end{figure}

\begin{figure}
\includegraphics[height=6truecm,width=7truecm]{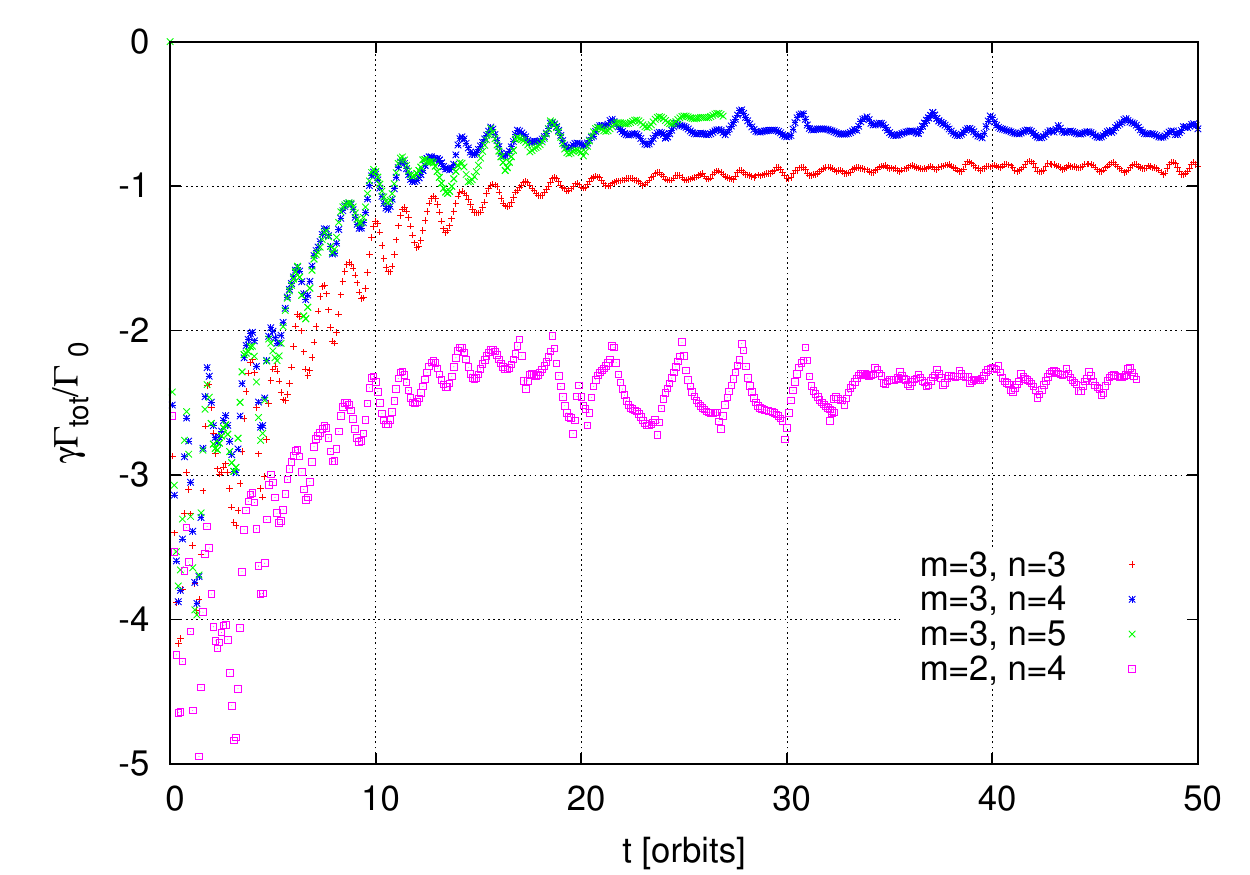}
\caption{Evolution of the total torque  with time for a planet of $3M_{\earth}$ and  three different
resolutions corresponding to $n=3$, $n=4$ and $n=5$ grid cells in the horseshoe region. Provided that the resolution is
at least of 4 cells in the horseshoe region the total torque becomes independent from the resolution. The case of a planet of $2M_{\earth}$ and  $n=4$ is also shown.}
\label{m3tt}
\end{figure}

A resolution corresponding to  $n=3$ grid cells in $x_{hs}$ is not enough
to properly compute the total torque: we found that 
 the value is underestimated for 
both the simulations with a planet of $5M_{\earth}$ (Fig.\ref{m5tt})  and a   $3M_{\earth}$ (Fig.\ref{m3tt}). For a $3M_{\earth}$  planet
  the steady state torque is more negative  for $n=3$ than for
  $n=4$  and $n=5$.
The choice of $n=4$  grid cells in the horseshoe appears to be a good compromise:
increasing the resolution further the total torque does not change any more, so that in the following we  show results obtained for   $n=4$.\par
For a planet of $2M_{\earth}$ (also shown in Fig.\ref{m3tt})
 the steady state torque is
negative and of the order of $-2.2\Gamma_0$.
Thus for the considered disc setting  (\citet{KBK09}) we can conclude that the transition from positive to negative torque occurs  for a planet of mass $m_p$: $3M_{\earth}<m_p<5M_{\earth}$.   \par

\subsection{Role of the potential}
\begin{figure}
\hskip -1 truecm
\includegraphics[height=6truecm,width=7truecm]{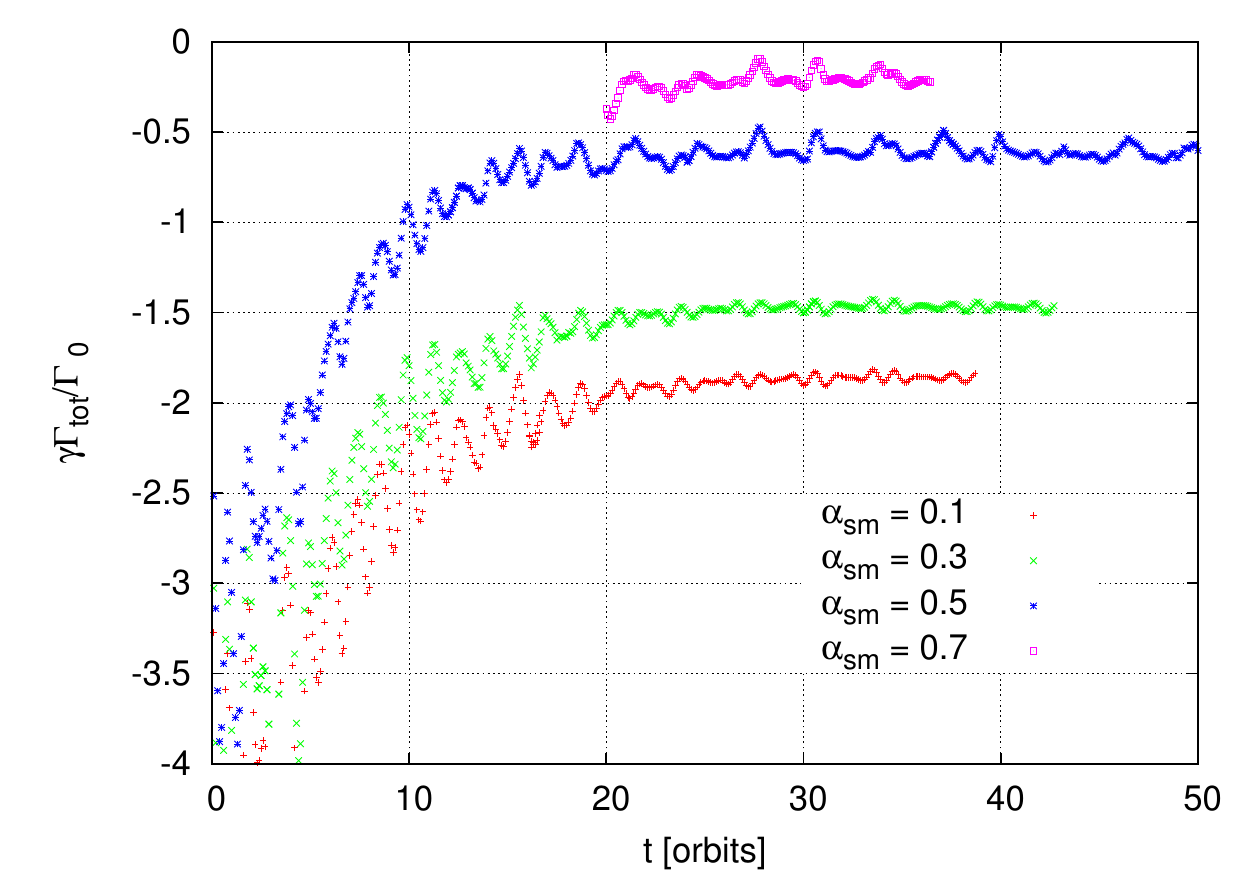}
\caption{Evolution of the total torque for a planet of $3M_{\earth}$ and a resolution of $n=4$  grid cells in the horseshoe region. Different simulations are run changing the value of
 the smoothing parameter $\alpha_{sm}$ in the cubic potential (Eq.\ref{cubic}).
Notice that the case $\alpha_{sm}=0.7$ has been restarted from the 
$\alpha_{sm}=0.5$ case at 20 orbits.
Deeper potentials (smaller values of  $\alpha_{sm}$)  give more negative  steady state torques.}
\label{alpha}
\end{figure}

In \citet{KBK09} (their Fig.14, top) it is shown that the value of the steady state torque depends on the planetary potential. Deeper  potentials  enhance the strength of the  torque which originates in the vicinity of the planet. 
For  a  $3M_{\earth}$ planet
 we have considered different values of the parameter 
$\alpha _{sm}$ in Eq.\ref{cubic}.  Fig.\ref{alpha} shows that the deeper the potential is, the more negative the total torque is.

Apparently, some effect in the close vicinity of the planet produce a negative
torque. 
In fact, we noticed that the steady state total torque for a $3M_{\earth}$ planet  with $n=4$  becomes positive if a Hill cut is applied.



\subsection{Comparison with analytic formula}
It is interesting to compare our results to those of
 \citet{Paaretal11} applying their formula to our disc parameters.
We recall that {it} captures the behavior of the torque caused by Lindblad resonances and the horseshoe
torque on low-mass planets embedded in 2D gaseous discs in the presence of viscous and thermal diffusion.
 The validity of this formula in the case of fully radiative 3D simulations has been checked in \citet{BK11}, showing that the formula predicts 
 a  torque that is  a factor  $3-4$ smaller than that 
observed in the simulations of planets in the range of $20M_{\earth}< m_P < 30M_{\earth}$.

\par
\begin{figure}
\includegraphics[height=6truecm,width=7truecm]{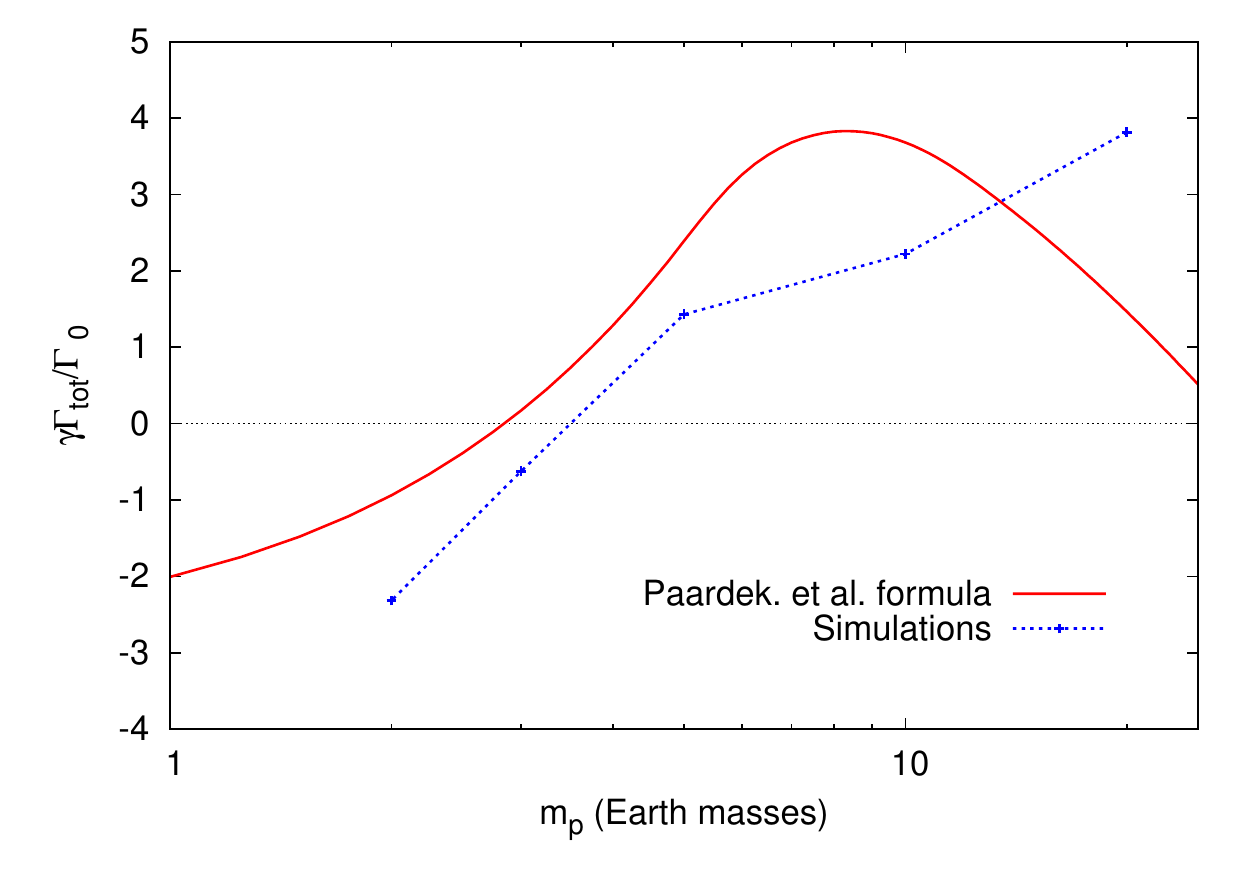}
\caption{Values of the steady state torque obtained with FARGOCA simulations
for planets of different masses. For all the cases the resolution corresponds to
$n=4$ grid cells in the horseshoe region
 and the smoothing parameter is $\alpha_{sm}=0.5$.  For comparison
the values of the torque provided by  \citet{Paaretal11} formula are plotted.
The values of the torque for a planet of $20M_{\earth}$ are presented in the
appendix Fig.\ref{ttm20}.}
\label{compa}
\end{figure}

In Fig.\ref{compa} we compare the results of our simulations with
the values provided by the \citet{Paaretal11} formula  at $r=5.2AU$ ($r=1$ in code units). 
We recall that in \citet{Paaretal11} formula the whole disc enters in the computation of the total torque so that no Hill cut should be applied in the numerical simulations as we did.
A trend towards more negative torque values appear in our data when decreasing the planetary mass. \par

\section{A fine description of the neighborhood of the planet}
\subsection{Radial torque}
\label{sect3}
\begin{figure}
\includegraphics[height=6truecm,width=7truecm]{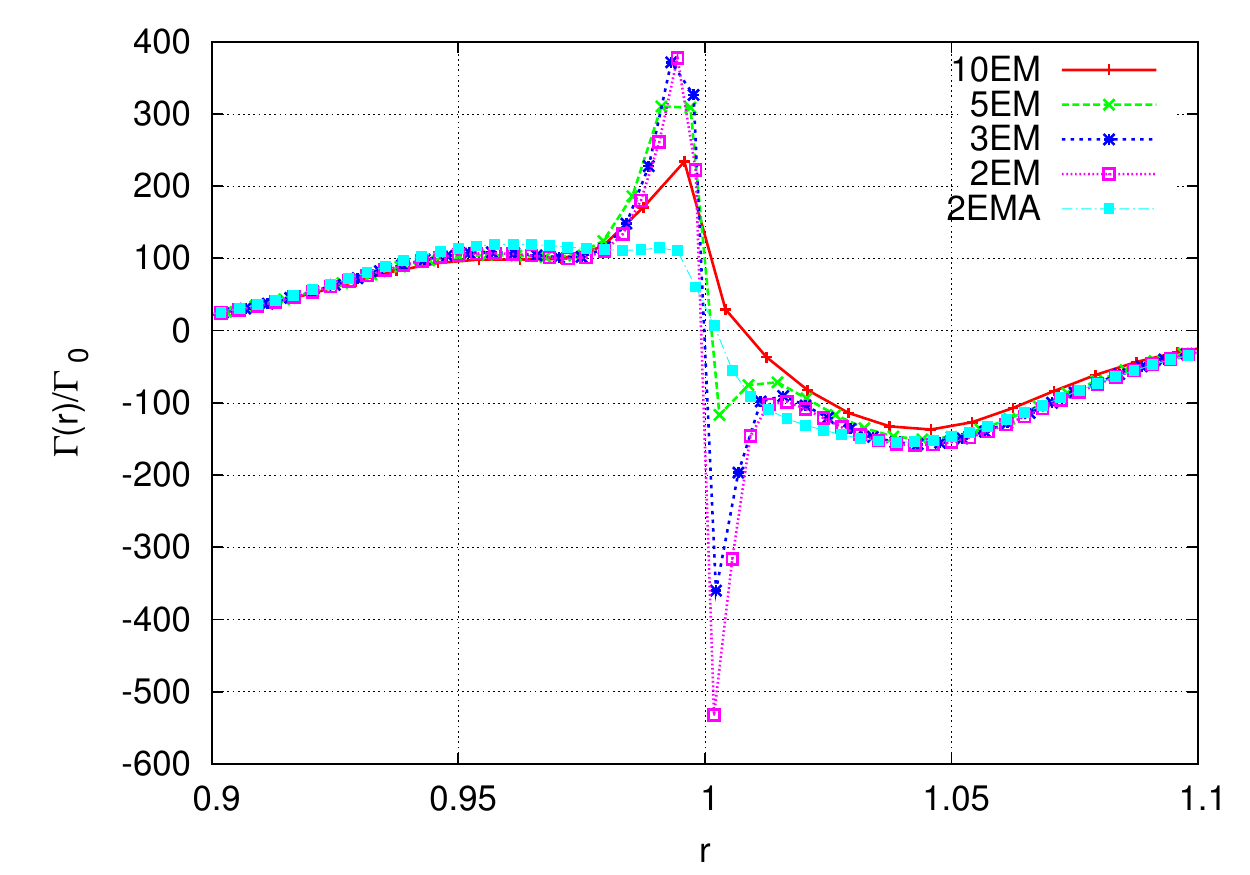}
\caption{ Radial torque density   normalized with respect to $\Gamma_0$.  The Lindblad torque scales with the square of the mass ratio $q$,  while unexpected contributions come from the close vicinity of the planets for $m_p<10M_{\earth}$. For comparison the radial torque density obtained for an adiabatic simulation is shown. }
\label{normtorque}
\end{figure}

In order to study the contribution to the  torque 
  in the close vicinity of the planets we show the radial torque density
$\Gamma (r)$. 
 The integral of $\Gamma(r)$ on the radial coordinate provides the total
torque $\Gamma_{tot}$ discussed in the previous section:
\begin{equation}
\Gamma_{tot}=\int _{r_{min}}^{r_{max}} \Gamma(r)dr
\end{equation}
 Fig.\ref{normtorque} shows the radial torque density  normalized with respect to $\Gamma_0$ ,  for the set of simulations shown in Table\ref{table:tab1} 
with a resolution corresponding to $n=4$.
We have also added the results obtained for a simulation with a $2M_{\earth}$
planet for the adiabatic case, i.e. considering in the r.h.s of  the energy equation (\ref{energy3D}) only the  the compressional heating term. The simulation is run with an
aspect ratio $H/r=0.037$  (i.e. the value obtained for the radiative equilibrium at $r=1$) and the same parameters as the case of a $2M_{\earth}$ planet
 for  $n=4$. 
The torque $\Gamma (r)$ for the adiabatic case is shown at a time of $40$ orbits
for which the total torque has reached a stationary value.
Different considerations can be done on Fig.\ref{normtorque}:
\begin{enumerate}
\item The Lindblad torque scales, as expected (\citet{GT80,Paaretal10})
 { with $\Gamma_0$, i.e.} with the square of the mass ratio $q$ .
\item 
Compared with the adiabatic case, where the component of the corotation torque
due to the temperature gradient saturates,
we notice that the total torque acting on a  $m_p= 10M_{\earth}$
has  a positive contribution just inside $r=1$;
also a small torque excess is observed  for $r>1$. Both features are expected
(\citet{PaardeMelle06}  in radiative discs as a result of the heating and cooling process of the gas librating in the horse-shoe region leading to a positive total torque

\item 
The corotation torque is expected to scale with $\Gamma_0$. 
At $r<1$ we observe a positive spike due to the  entropy related part of the corotation torque (non linear contribution or horseshoe drag).
Decreasing the planetary mass,  we expect that the horseshoe  drag tend towards the linear corotation torque.  Instead, in Fig.\ref{normtorque} we observe that the positive contribution remains when decreasing the planetary mass. 
\item  A new and totally unexpected feature appears for $r>1$. 
We observe a negative spike which is not seen for  $m_p \geq 10M_{\earth}$
and which becomes more and more important when decreasing the mass of the planet.
\item The positive and negative spikes contributions appear to be asymmetric,
the negative spike providing a larger contribution which is responsible 
for the negative total torque, i.e. for the transition from outward to inward migration.
\end{enumerate}

\begin{figure}
\includegraphics[height=6truecm,width=7truecm]{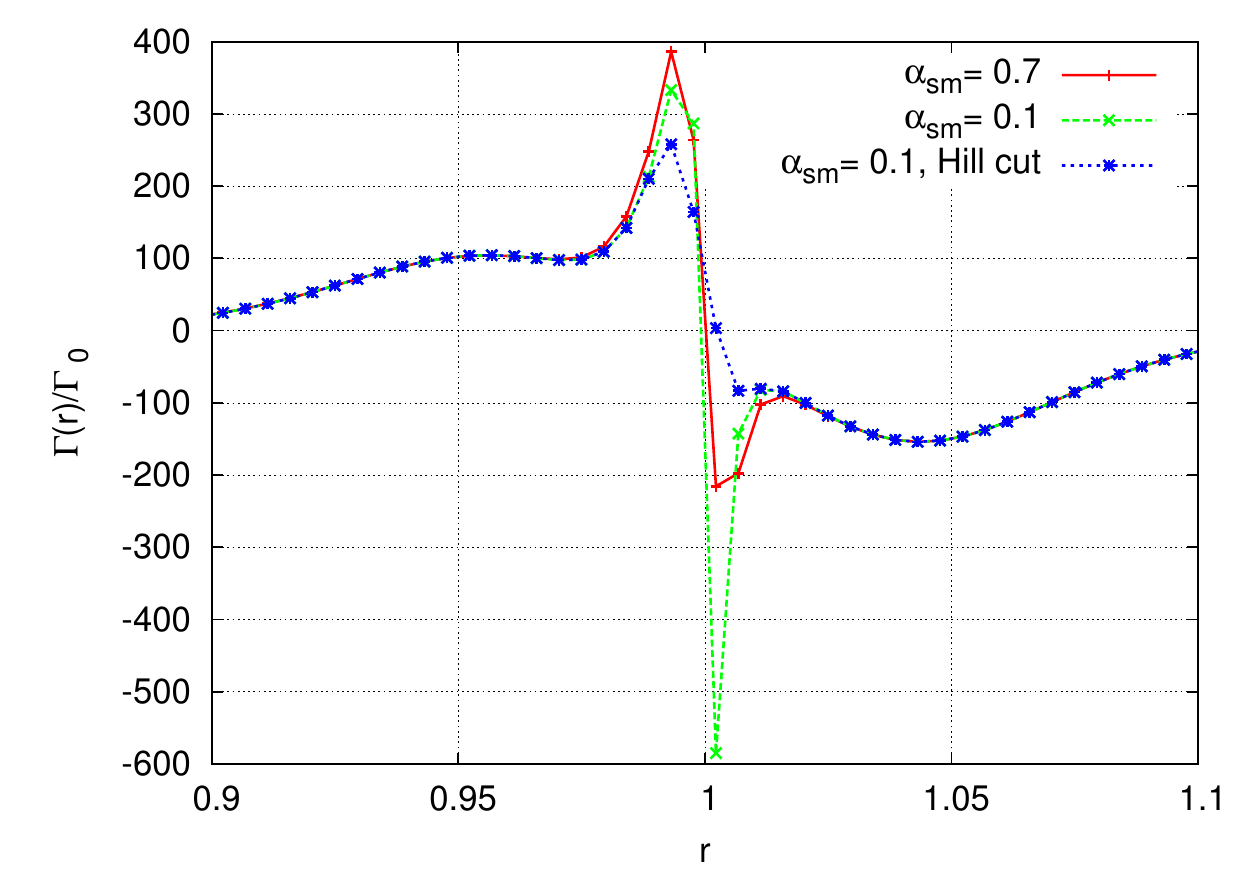}
\caption{ Radial torque density for a planet of $3 M_{\earth}$ and
resolution $n=4$, using the cubic
 potential (Eq.\ref{cubic}) with two different values of the smoothing parameter: $\alpha_{sm}=0.1$ and $\alpha_{sm}=0.7$. The contribution of the negative spike observed for $r >1$ is more important for deeper potentials. For comparison the case
$\alpha_{sm}=0.1$ is plotted considering the Hill cut.}
\label{spikealpha}
\end{figure}

At the light of this picture it is easy to see that the exclusion of the inner part of the Hill sphere of the planet in the torque computation  
would reduce mainly the effect of the negative spike which is  closer to the planet
 location than the positive spike, which will then ``survive'' the Hill-cut.
 This is shown in Fig.\ref{spikealpha} where the blue curve is similar to the
green one except that a Hill-cut has been applied.
 This fact  explains the change of sign of the total torque for
the $3M_{\earth}$ case mentioned before. Moreover, the negative spike appears to be  more important for deeper  potentials (Fig.\ref{spikealpha}) thus explaining results of Fig.\ref{alpha}. \par 

\subsection{Local dynamics}
In order to understand the origin of these torques contributions
together with their asymmetry we investigate the dynamics in the vicinity of the planet.
\par
\begin{figure}
\includegraphics[height=7truecm,width=7truecm]{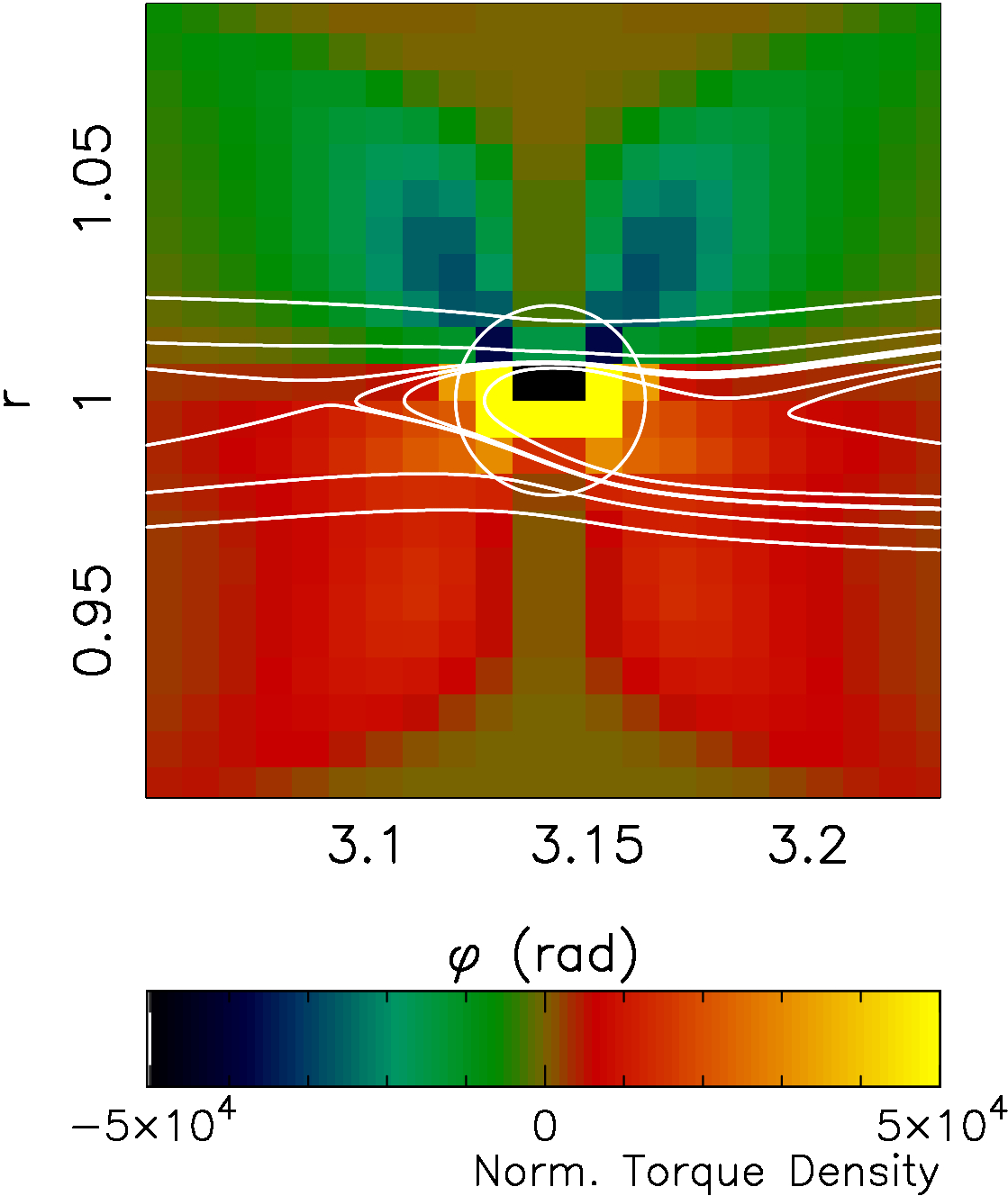}
\caption{Normalized torque density $\tilde \Gamma_{\theta_p}$ acting on a planet of $10 M_{\earth}$ caused by the density in each individual midplane grid-cell as defined in Eq.\ref{symme}. { Some streamlines are shown by white curves. The circle around the point $(\pi,1)$ shows the Hill's sphere of the planet.} }
\label{torque10}
\end{figure}
\begin{figure}
\includegraphics[height=7truecm,width=7truecm]{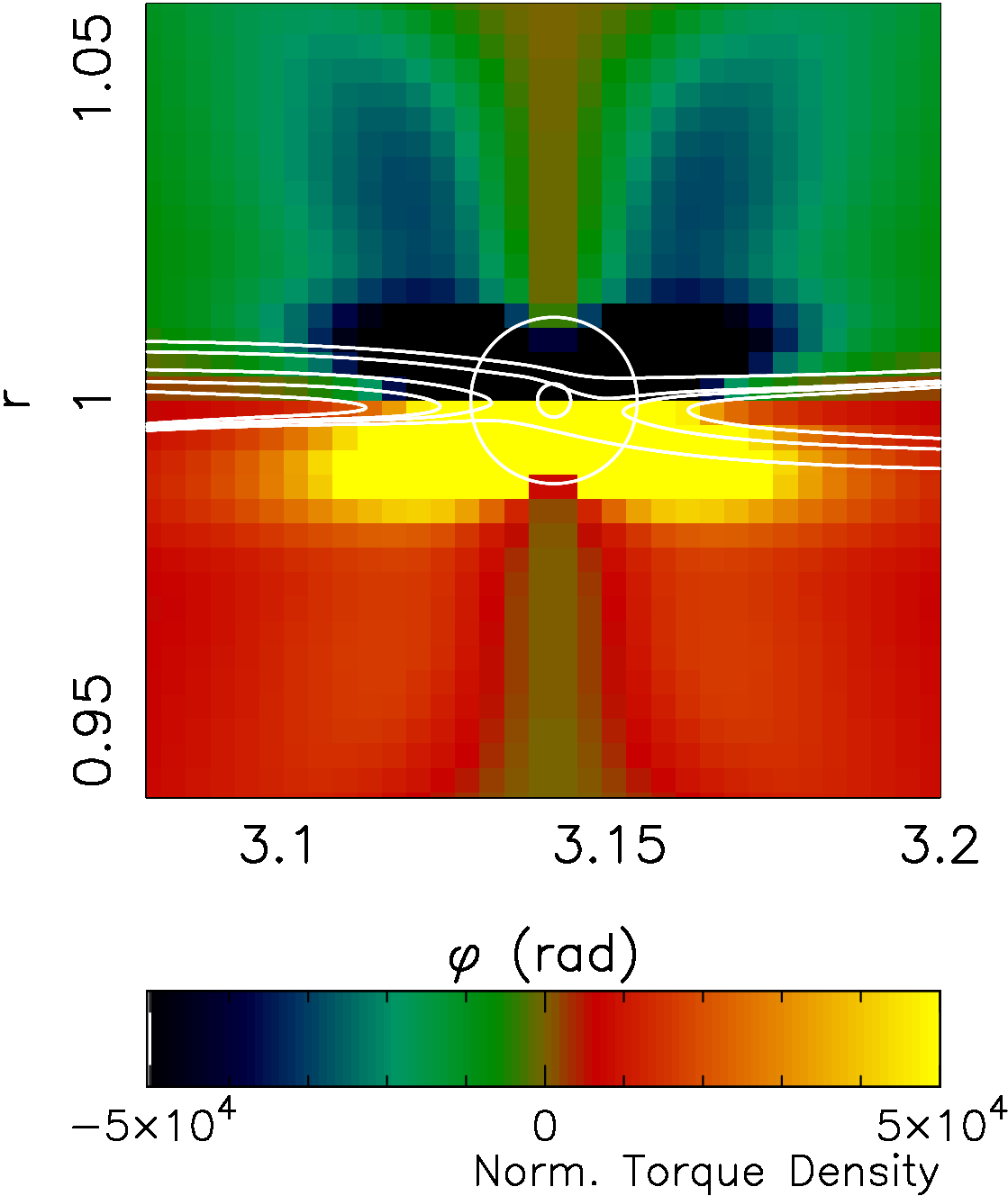}
\caption{Same as Fig.\ref{torque10} for a planet of $2 M_{\earth}$. The small circle around the point $(\pi,1)$ shows the Bondi's sphere of the planet.}
\label{torque2}
\end{figure}

Fig.\ref{torque10} and \ref{torque2} show
 the contributions of individual midplane grid-cells to the torque on the planets for respectively a $10M_{\earth}$ and a $2M_{\earth}$ planet. The torque is constructed by adding to 
each grid-cell the contribution of of the symmetric cell with respect to
the planet location; for a planet in $r_p,\theta_p,\varphi _p$ we have on a coordinate
 $r,\varphi$:
\begin{equation}
\tilde \Gamma_{\theta_p}(r,\varphi) =  \Gamma_{\theta_p}  (r,\varphi)+  \Gamma_{\theta_p}  (r,2\pi-\varphi) 
\label{symme}
\end{equation}
{ where $\Gamma_{\theta_p}  (r,\varphi)$  is a volume torque density,
 i.e. it is
the torque exerted on the planet 
by the  mass density  $\rho$
 placed at $(r,\varphi,\theta_p)$.
The torque density has been normalized with respect to $\Gamma_0$.}\par
The quantity $\tilde \Gamma$ is symmetric with respect to $\varphi_p$, so that only the left side of Fig.\ref{torque10} and \ref{torque2}
($\varphi <\varphi_p$) can be taken into account to estimate global effects.
A positive (negative) value of 
$\tilde \Gamma_{\theta_p}(r,\varphi)$ in the whole interval
$0<\varphi<\pi$ corresponds to positive (negative) radial
density torque on the  midplane.
We notice that the change of sign of $\tilde \Gamma_{\theta_p}(r,\varphi)$
in the vicinity of the planet location that we observe 
in  Fig.\ref{torque10} and \ref{torque2} 
 is in agreement 
with  Fig.\ref{normtorque} where the radial torque density has been
obtained considering also the contribution of the grid-cells in the vertical 
direction and not only on the midplane as given by Eq.\ref{symme}.

Therefore it is justified to search for the origin of the negative spike 
 at $r>1$ analyzing the values of different gas quantities only on the midplane.
At first we notice in Fig.\ref{torque2} 
strong  negative (positive) values of $\tilde\Gamma_{\theta_p} (r,\varphi)$ 
at $r>1$ ($r<1$)  in the vicinity of the planet location that are not observed in Fig.\ref{torque10}. \par

 The circle around the point $(\pi,1)$ shows the Hill's sphere of the planet. 
{ In Fig.\ref{torque2} we have also plotted the Bondi's sphere of the
planet, with  Bondi radius defined by:
\begin{equation}
R_B= {Gm_p \over c_s^2}
\label{bondirad}
\end{equation}
It appears clearly that the region concerned by the positive and negative 
torque excesses  is larger than the Bondi's sphere.  
 From the  streamlines over plotted we can observe that   many  streamlines pass in the vicinity of the planet outside of the Bondi's sphere.   } 


\subsection{Density analysis.}
\begin{figure}
\includegraphics[height=6truecm,width=6truecm]{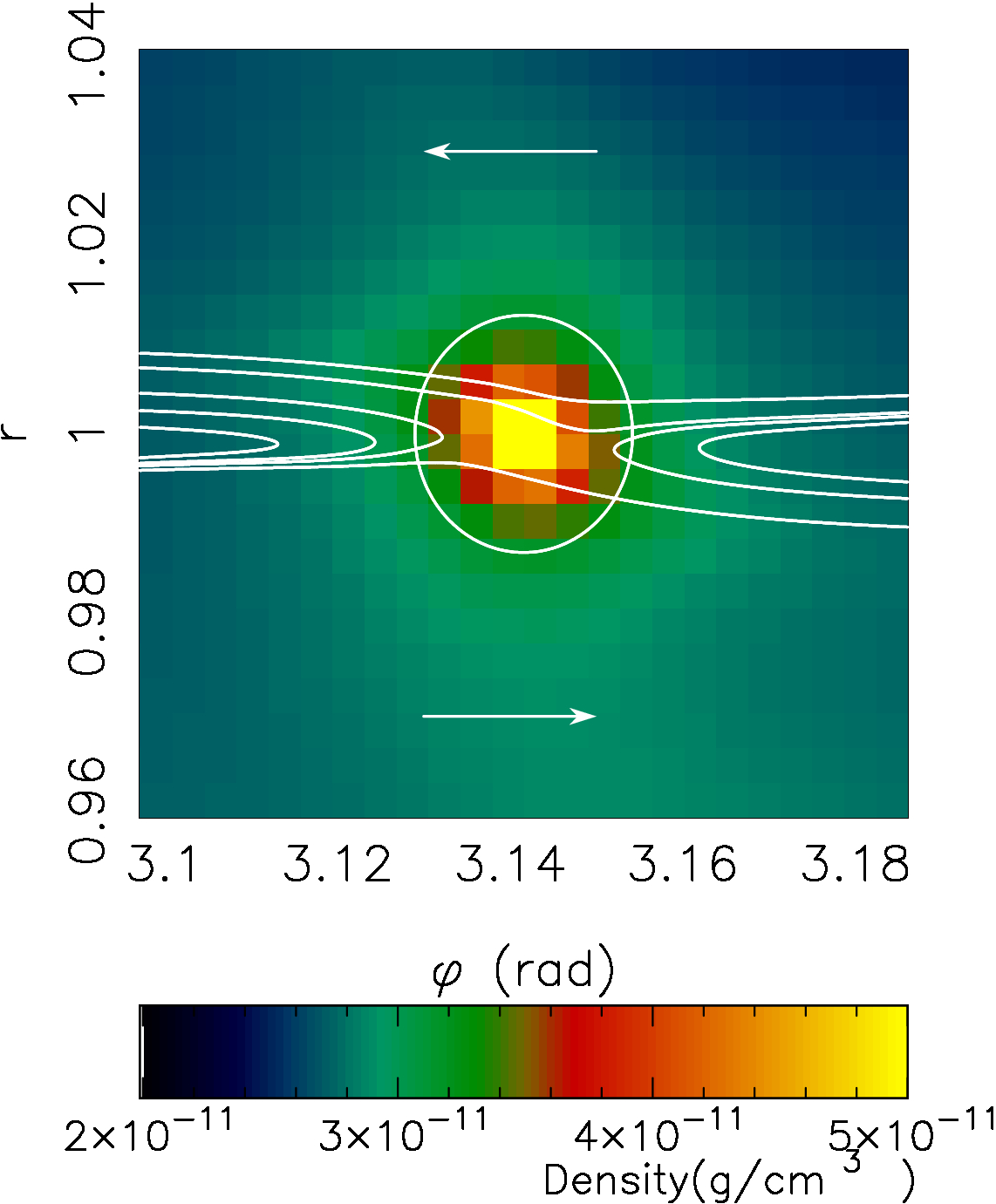}
\includegraphics[height=6truecm,width=6truecm]{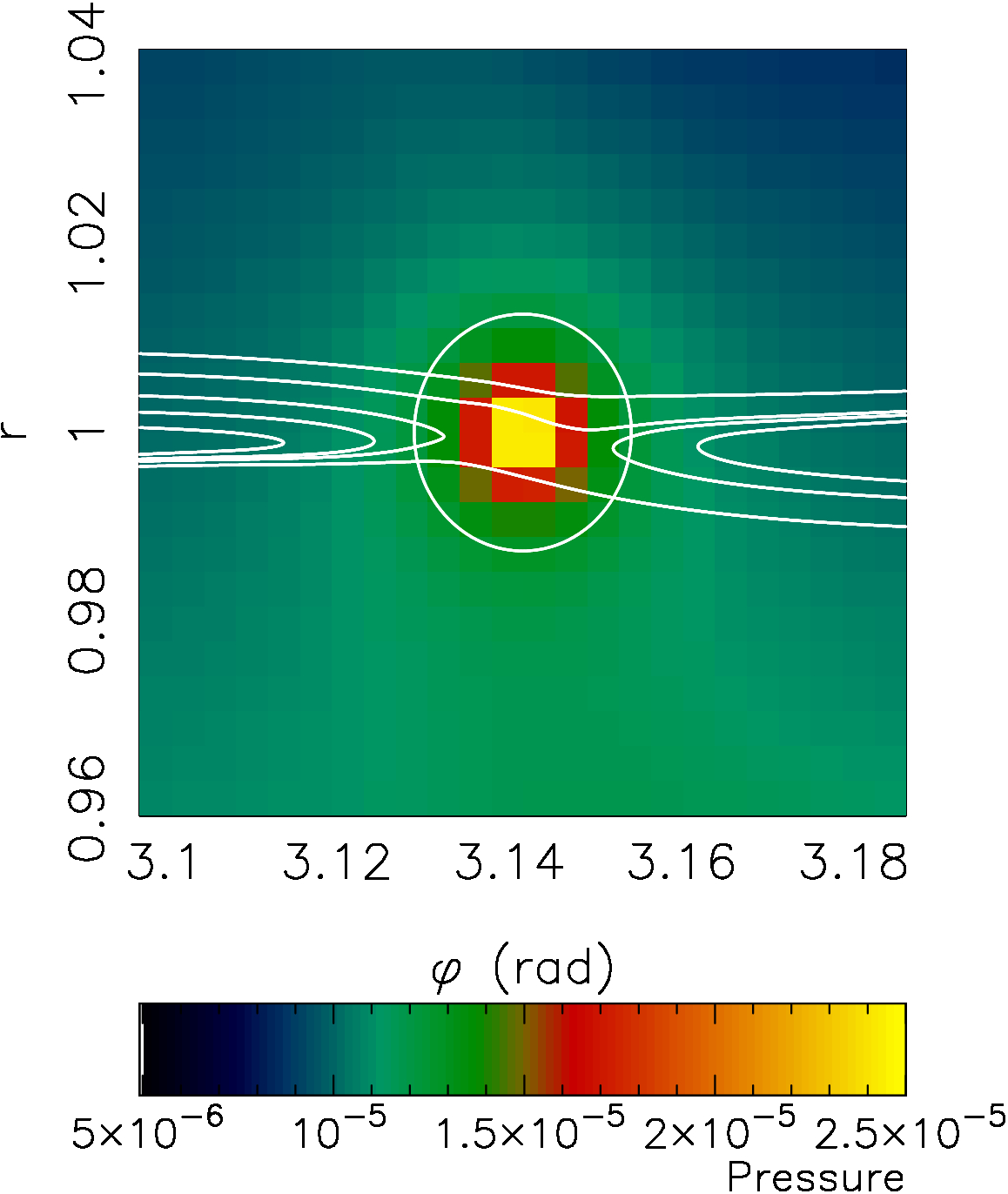}
\includegraphics[height=5.4truecm,width=5.4truecm]{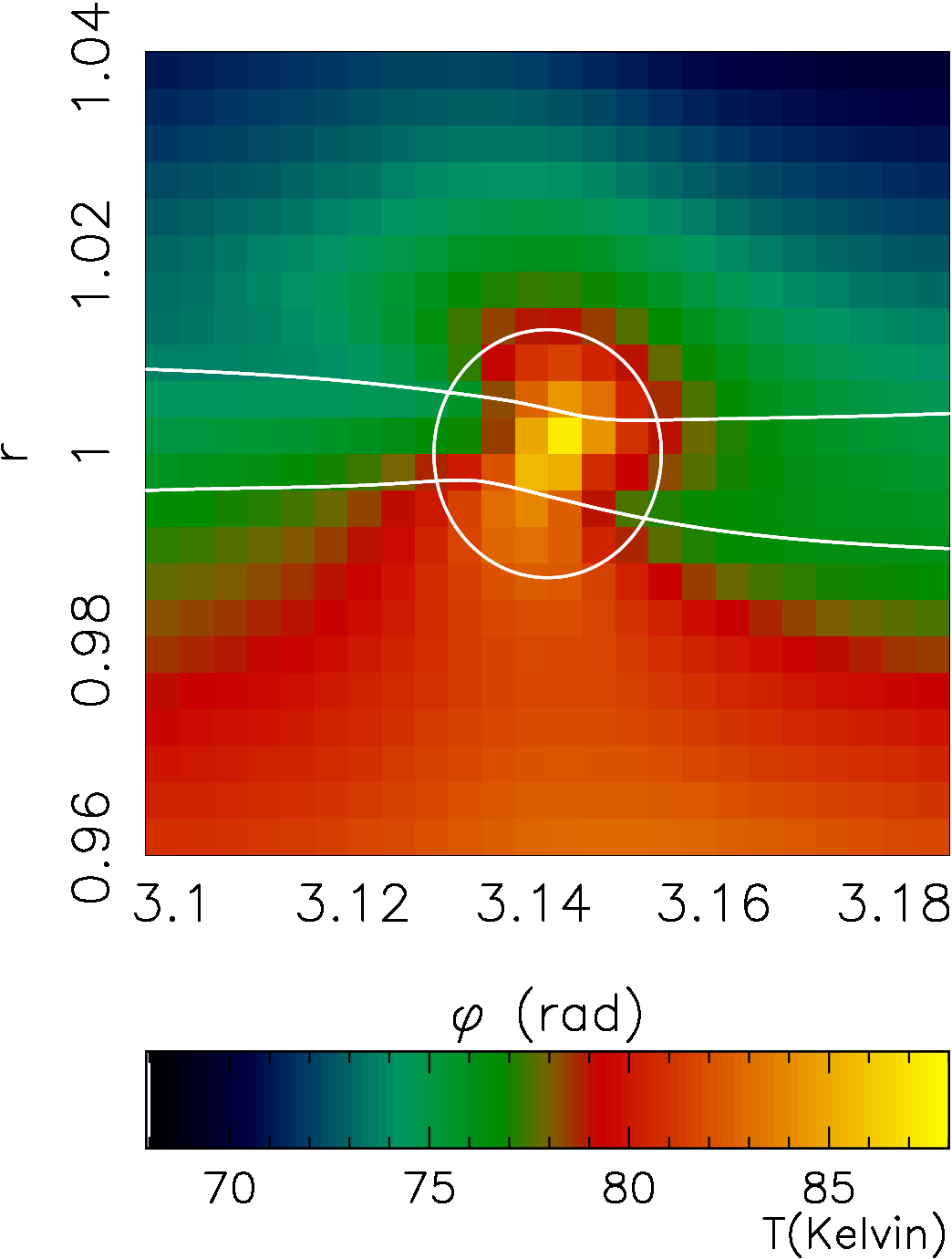}
\caption{Midplane plots for  a $2M_{\earth}$ planet. 
{\bf Top panel}:  density field, {\bf middle panel}: pressure field,  {\bf bottom panel}:  temperature field.
Some streamlines are shown by white curves. The arrows in the top panel indicate the flow circulation:  from right to left  for $r>1$ and  from left to right  for $r<1$.
Two streamline are shown in the bottom panel
passing through points symmetric  with respect to the planet position:
$r-r_p= 5.5\,10^{-3}, \varphi-\varphi_p=-5.5\, 10^{-3}$ (top), $r-r_p= -5.5\,10^{-3}, \varphi-\varphi_p=5.5\, 10^{-3}$ (bottom).}
\label{m2D}
\end{figure}

\begin{figure}
\includegraphics[height=7truecm,width=7truecm]{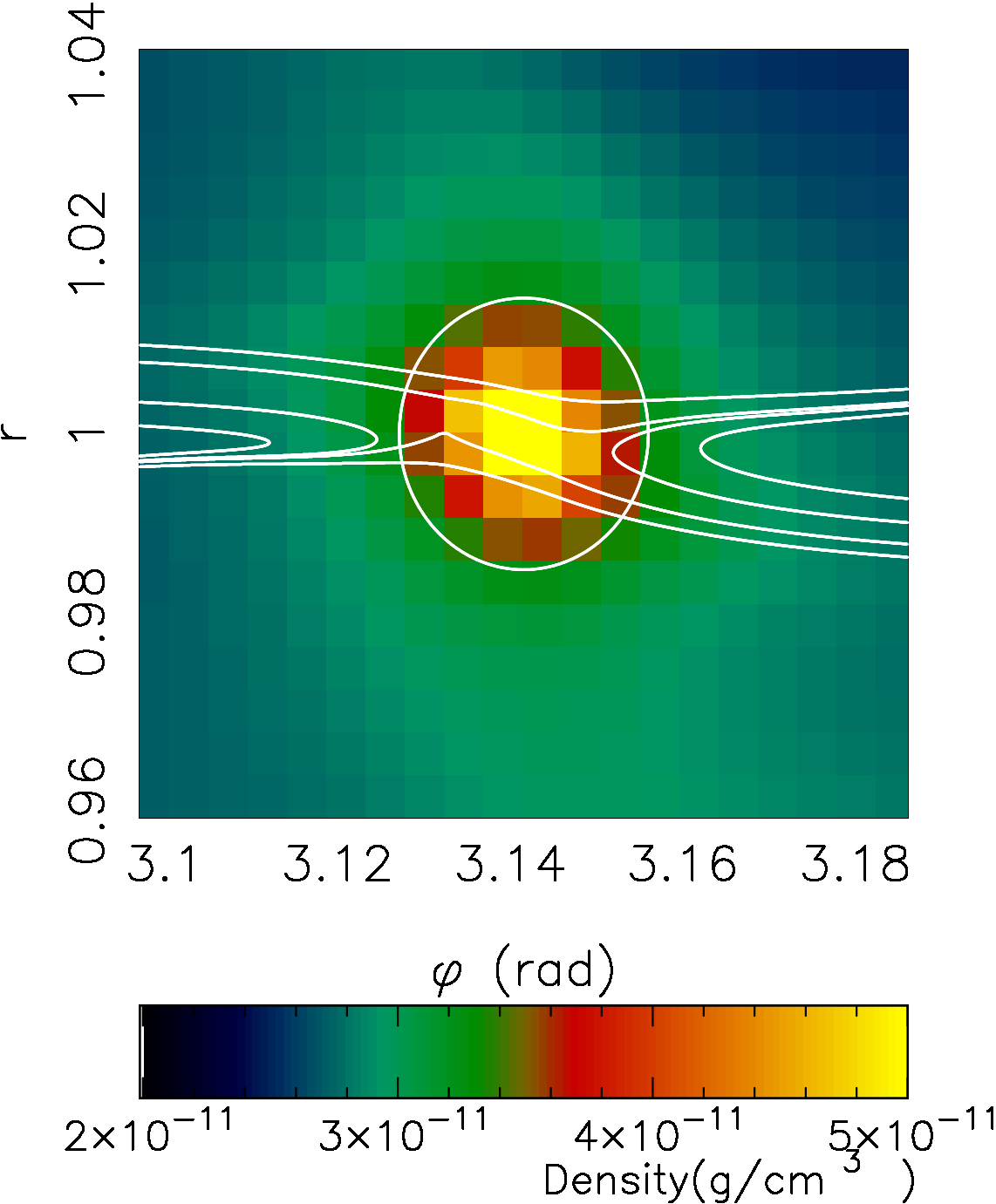}
\caption{Midplane density field for a $3M_{\earth}$ planet (resolution $n=4$,
$\alpha_{sm}=0.5$).}
\label{m3D}
\end{figure}

\begin{figure}
\includegraphics[height=7truecm,width=7truecm]{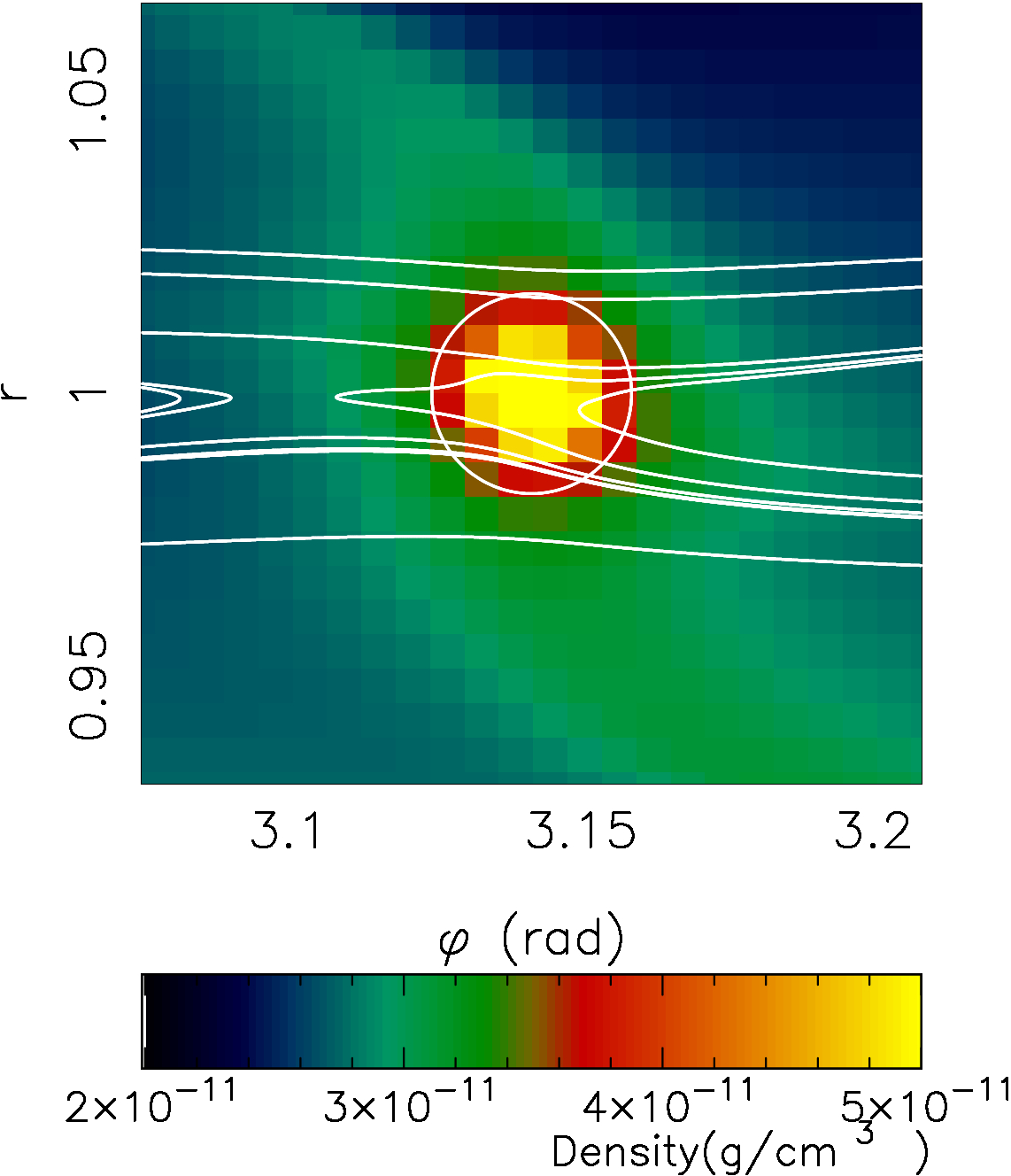}
\caption{Midplane density field for a $5M_{\earth}$ planet (resolution $n=4$,
$\alpha_{sm}=0.5$). }
\label{m5D}
\end{figure}

Looking carefully at the density plot  (Fig.\ref{m2D}, top panel,  midplane values are shown)
we see that the density distribution around a $2M_{\earth}$  planet is asymmetric.
In particular, just outside the planet orbit  ($r>1$) 
the density is larger  at $\varphi < \varphi_p$ than symmetrically at  $\varphi > \varphi_p$. This gives the negative contribution to the total torque in Fig.\ref{torque2}. Instead,  just inside the planet orbit  ($r<1$) the density is larger  at $\varphi > \varphi_p$ than at  $\varphi < \varphi_p$.  This gives the positive  contribution to the total torque in Fig.\ref{torque2}. The same is observed for higher planet's masses (Fig.\ref{m3D},\ref{m5D}). \par
We remark that, because
 the gas circulates from right to left  for $r>1$ and 
 from left to right  for $r<1$, both density enhancements occur downstream
relative to the encounter of the gas with the planet.

 \par
On Fig.\ref{m2D}, middle panel, we see that the vicinity of the planet is a region of high pressure and that the pressure field is azimuthally perfectly
symmetric. As a consequence, the equation of state implies that variations
of density and temperature with respect to symmetry are anticorrelated
{ (see Fig.\ref{m2D}, bottom panel)}.

We explain what is observed as follows:
when the gas goes through the high density and high pressure region close to the planet it is compressed and its temperature increases (adiabatic heating). If the equation of state were adiabatic the temperature,
density and pressure would come back to the original values  after leaving the planet vicinity. But, when radiative transfer is included,
the gas cools during the entire  encounter with the planet, i.e. when it is hotter than the unperturbed temperature.  Consequently, as it comes back to the original pressure, the gas is colder after the encounter with the planet  than it was before the encounter, because it has lost internal heat. 
This leads to the formation of a cold and therefore dense finger downstream
 relative to the planet location, namely ahead of the planet just inside of its orbit, and behind the planet just outside of its orbit.\par
\begin{figure}
\includegraphics[height=7truecm,width=7truecm]{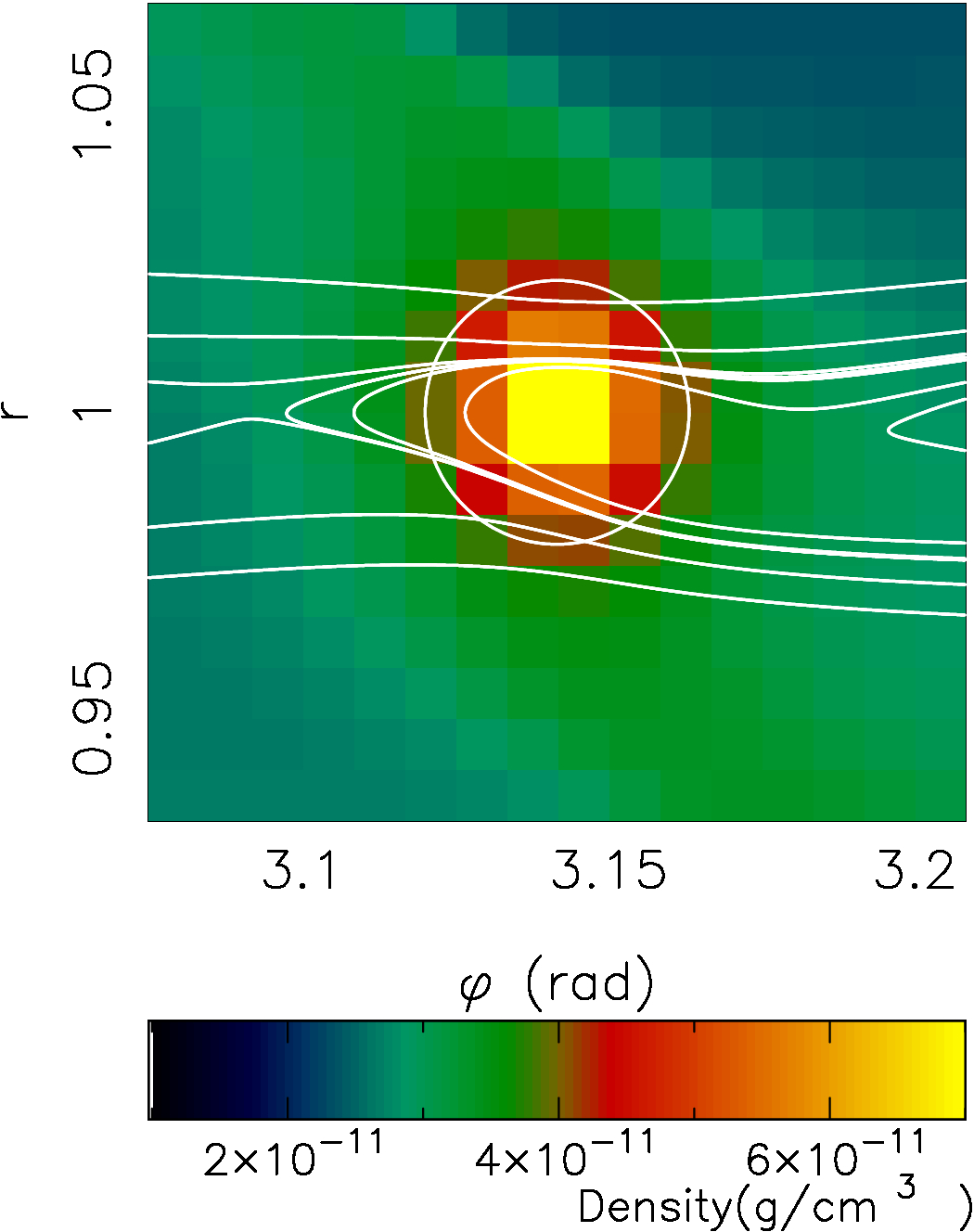}
\caption{Midplane density field for a $10M_{\earth}$ planet (resolution $n=4$,
$\alpha_{sm}=0.5$).   }
\label{m10D}
\end{figure}

In the case of a $5M_{\earth}$ planet (Fig.\ref{m5D}) we notice  only one stagnation point at $\varphi <\varphi_p$
at $r$ slightly smaller than 1. Therefore, also some libration streamlines
pass through the high density and high pressure region and provide, through
the same mechanism, an excess of density  after the encounter with the planet for $\varphi >\varphi_p$, $r<1$. The same occurs for the planet of $10M_{\earth}$
(Fig.\ref{m10D}).  Therefore, the positive contribution to
the torque at $r<1$ seems to be due to two effects:  the first one is due 
to the libration and circulation streamlines passing through
the high pressure region as we just described. The second effect
 is due   to horseshoe streamlines that pass further from the planet and
 do not heat up significantly during the encounter, thus bringing gas from a colder to a hotter part of the disc 
as described by \citet{PaardeMelle06}. 

 Moreover, the distribution of the gas density around the planet observed at $10M_{\earth}$  appears to be more symmetric than for smaller masses.
We notice that, although some circulation streamlines pass through the high
density region, their distance from the planet is larger than for smaller masses. As we will show in the next section the effect of adiabatic heating and
radiative cooling and therefore the
downstream density enhancement vanishes  with increasing distance from the planet.

\par At our knowledge  this phenomenon was not observed in previous 2D 
simulations probably because of the smoothed potential (Eq.\ref{smooth})
which is used in 2D simulations to soften the effect of the vertical gas column
as explained in the introduction.
In the following section we show results obtained with 2D simulations 
in which we use the cubic potential and a  simple cooling prescription in which the cooling time appears as a parameter. The aim is to recover the proposed mechanism and to study its characteristics as a function of the cooling properties of the disc.

\section{2D simulations}

In order to better understand the basic physics at play around a small mass planet, we have performed 2D simulations. The potential of the planet was smoothed like in 3D simulations, given by Eq.~(\ref{cubic})\,; by doing so, we do not pretend to model accurately a thick disc, but to study the dynamics in the midplane.

Having only two dimensions allows a higher resolution, but the radiative cooling can not be modelled accurately. This turns out to be an advantage, as we can use a simplified, controlled treatment of the heating/cooling processes. We use the following energy equation\,:
\begin{equation}
\label{energy2D}
{\partial E \over \partial t} + \nabla \cdot (E\vec v)
 =  -p\nabla \cdot \vec v  - c_v\Sigma \frac{T-T_0}{\tau_c}
\end{equation}
where $\Sigma=\int\rho\,{\rm d}z$ is the surface density, $\tau_c$ is the cooling time, and $T_0$ is the initial temperature, defined as $T_0(r)=(GM_*/r)h^2$. In short, this is an adiabatic EOS, with exponential damping of the temperature perturbations. { Notice that we do not consider the viscous heating term as in Eq.\ref{energy3D}  since the main source of heating close to the planet is the compressional heating.}

\begin{figure}
\includegraphics[height=6truecm,width=6truecm]{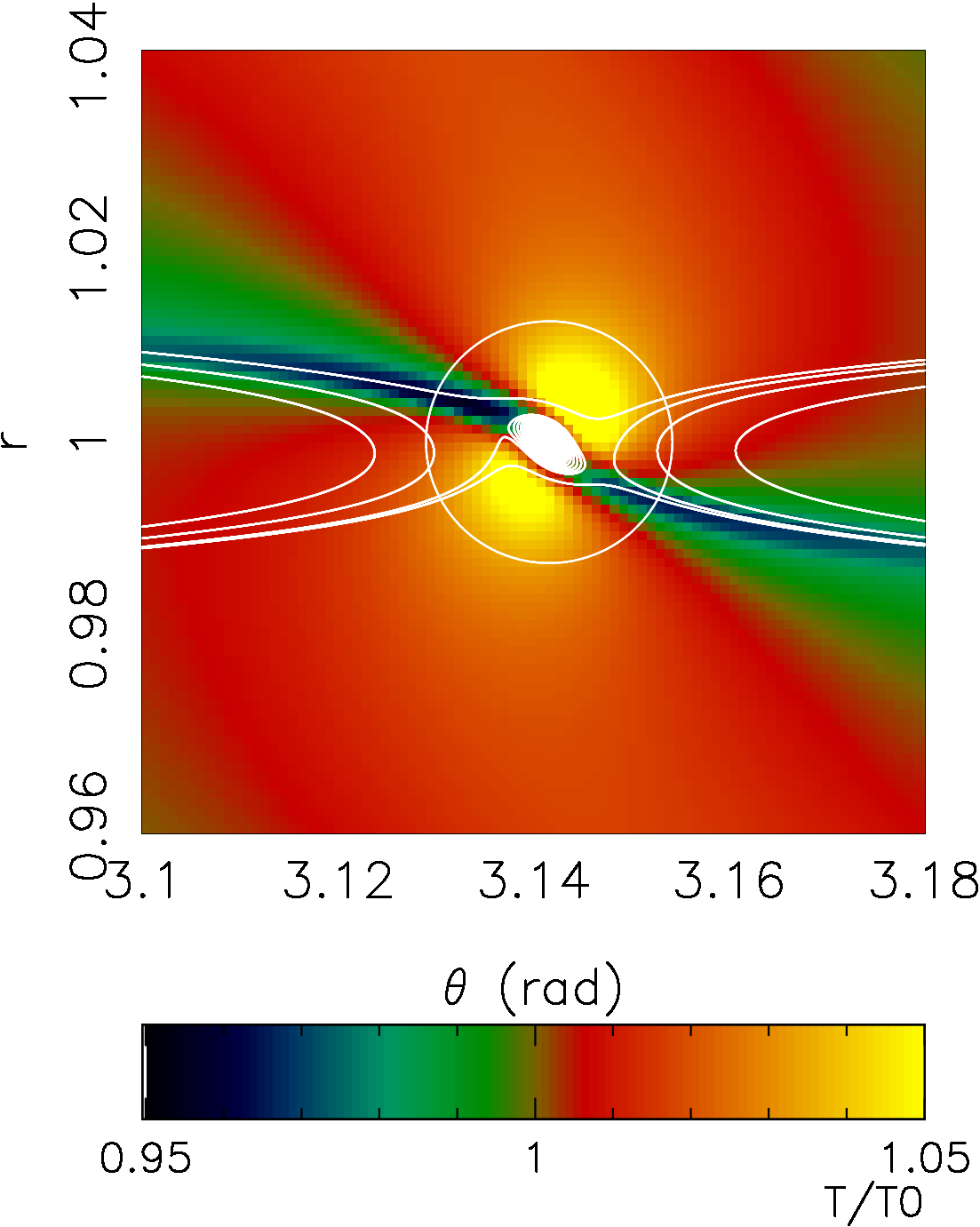}
\includegraphics[height=6truecm,width=6truecm]{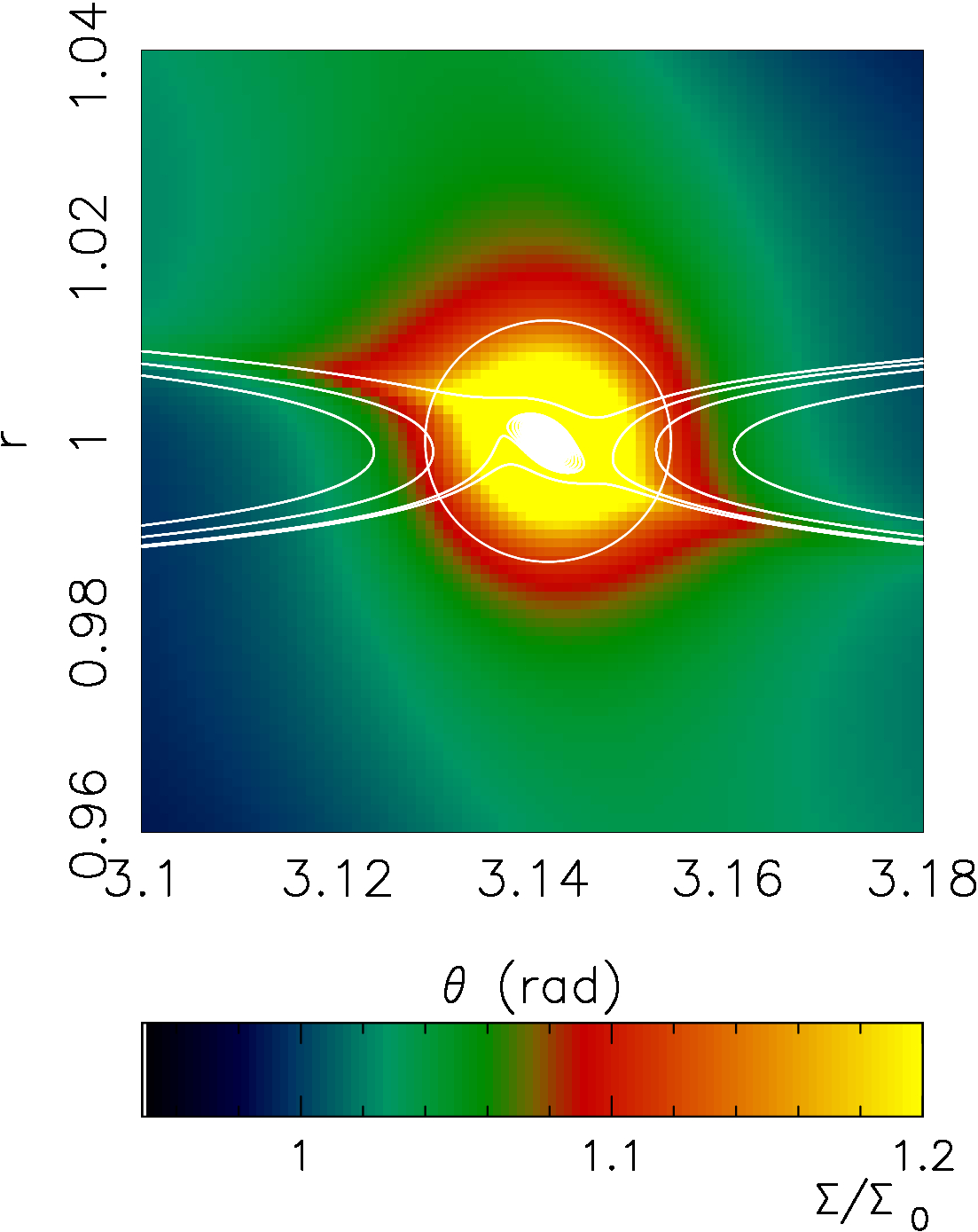}
\includegraphics[height=6truecm,width=6truecm]{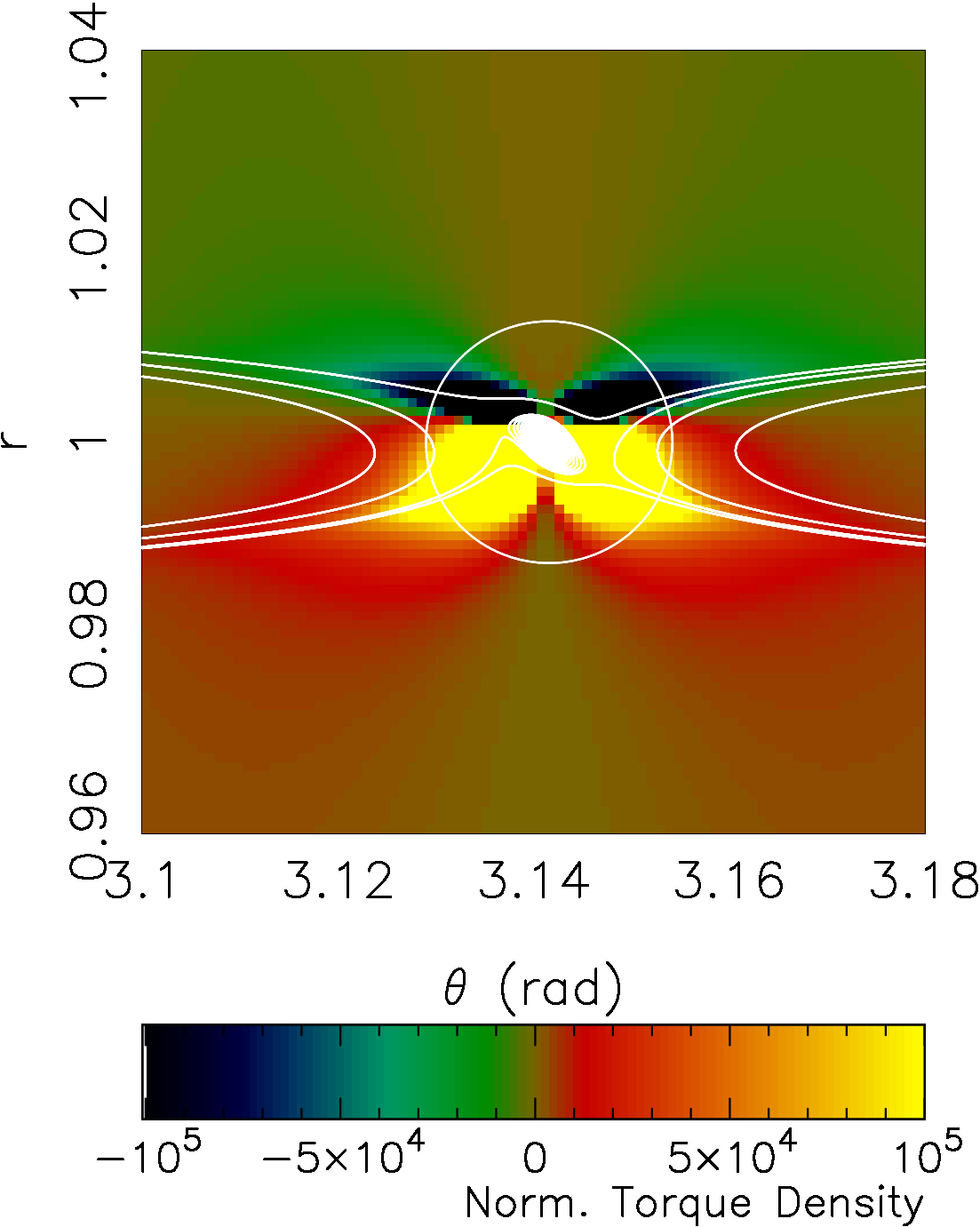}
\caption{{\bf 2D Simulation}: perturbed temperature ({\bf top panel}),  perturbed  density ({\bf middle panel}) and normalized torque density ({\bf bottom panel})  around a $2 M_\oplus$ planet after 60 orbits. The energy equation is given by Eq.~(\ref{energy2D}) with $\tau_c=\sim 2/\Omega_p$. About half a cooling time after the maximum of the compression (at conjunction with the planet), the gas becomes cooler than initially, making a diagonal cold finger (in blue, middle panel). { The cold  finger of gas corresponds to a denser diagonal finger. The comparison with the 3D simulations results of Fig.\ref{torque2} and \ref{m2D}   shows that the same mechanism acts both the 2D and  3D case. }}
\label{fig:adc2}
\end{figure}

We have done two sets of  simulations on a computational domain $(r,\theta)$  consisting of an annulus of the protoplanetary disc extending from $r_{min}$ to $r_{max}$
with $r_{min}= 0.6$ , $r_{max}=1.65185$ on a grid of  $N_r\times N_\theta =568\times 3392$ gridcells for set $A$;   $r_{min}= 0.8$ , $r_{max}=1.325925$ on a grid of  $N_r\times N_\theta =568 \times 6684$ gridcells for set $B$. { Both sets of simulations have squared gridcells of side $1.85\,10^{-3}$ (set A) and
$9.25\,10^{-4}$ (set B) at the planet location, $h=0.05$ and a constant viscosity coefficient of $10^{-5}$ in code units ({or $10^{11}\rm{m^2/s}$=}$10^{15} \rm{cm^2/s}$)}.
Figure~\ref{fig:adc2}, top panel, shows (set $B$) the perturbed temperature   $T/T_0$ in the neighborhood of a 2 Earth mass planet after 60 orbits, with $\tau_c=(2/\Omega_p)(r/r_p)^2$. Inside the Hill sphere of the planet, compressional heating takes place. However, along the outgoing streamlines, the temperature is significantly colder than before, in agreement with what was seen in 3D. This happens because the expansion cools a gas that has already lost some internal energy due to the radiative cooling. The temperature then slowly comes back to the unperturbed one (following the prescribed exponential damping).
{ The colder finger can be traced as well in the density plot  (Figure~\ref{fig:adc2}, middle panel), through a high density finger which gives a
torque contribution as shown in Figure~\ref{fig:adc2}, bottom panel.}

\begin{figure}
\includegraphics[height=5truecm,width=5.5truecm]{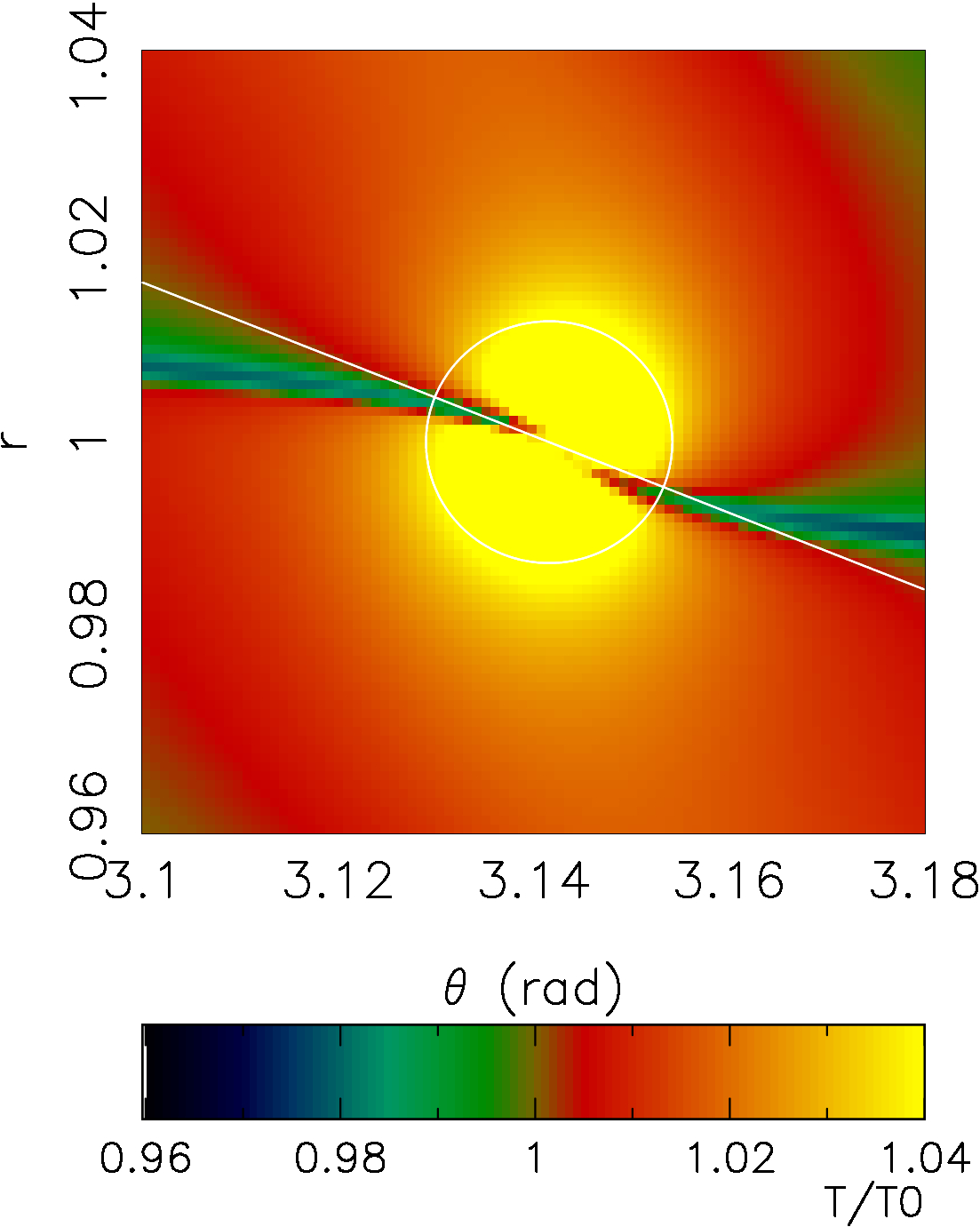}
\includegraphics[height=5truecm,width=5.5truecm]{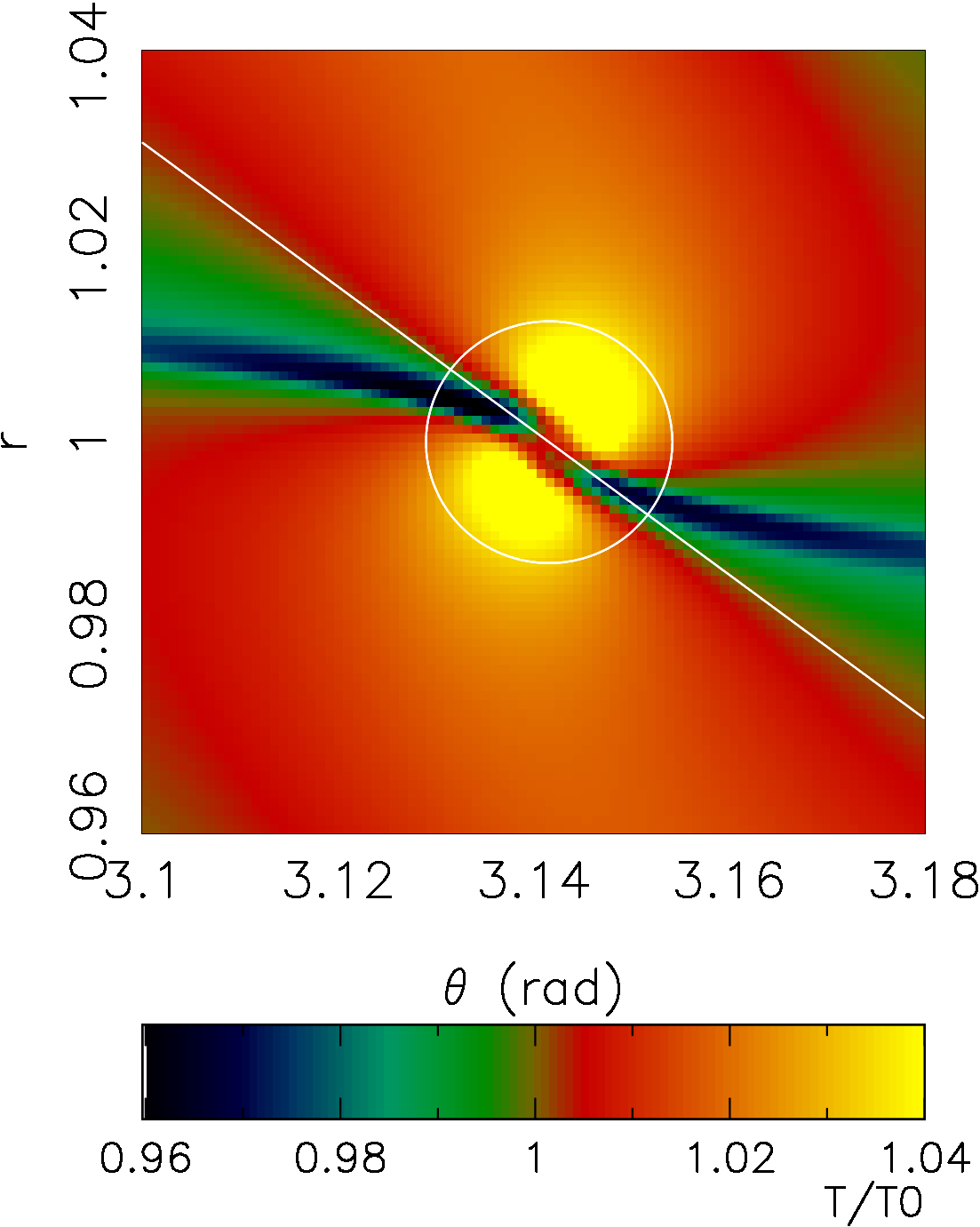}
\includegraphics[height=5.5truecm,width=5.5truecm]{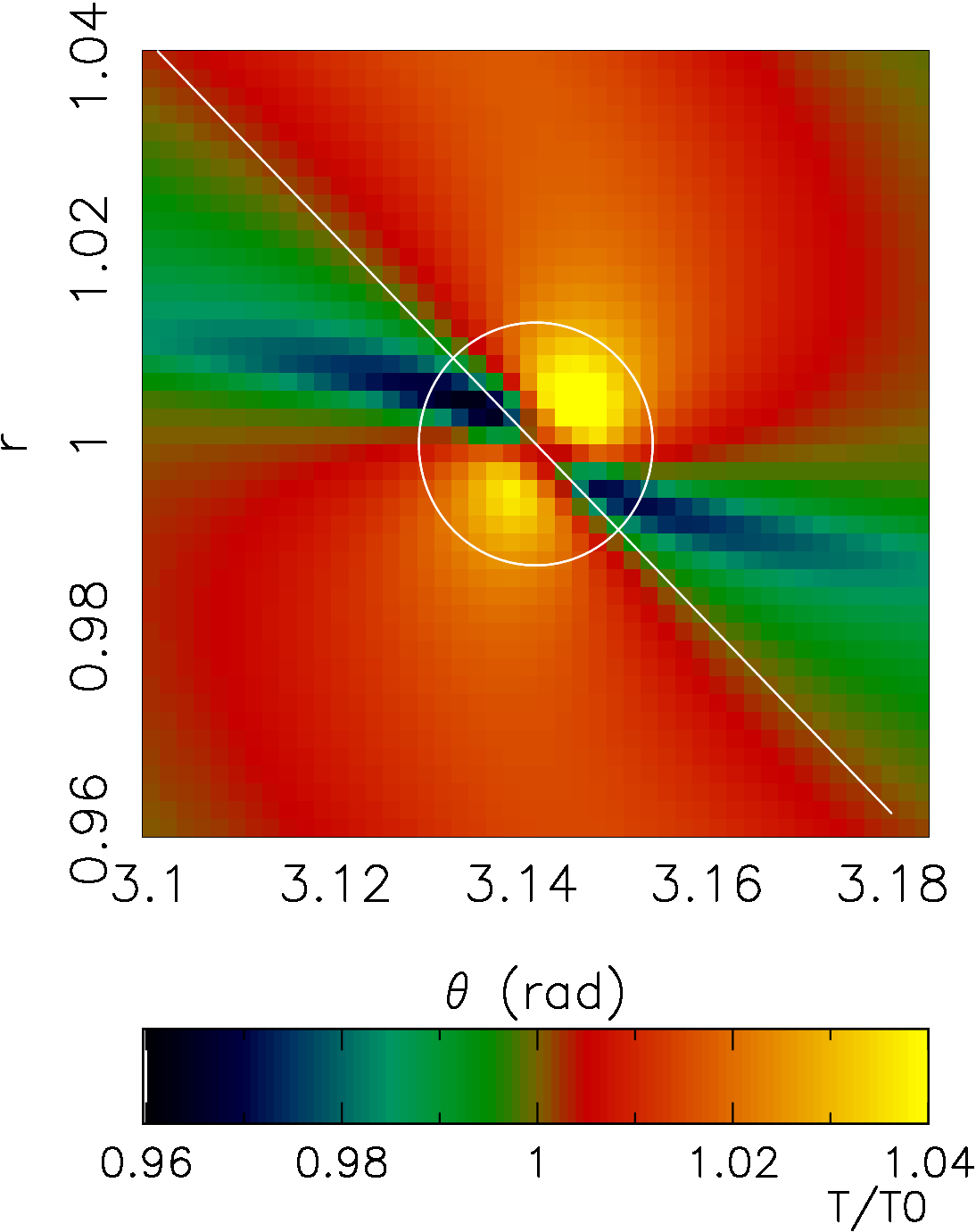}
\caption{{\bf 2D Simulation}: perturbed temperature variations around a $2 M_\oplus$ planet after 10 orbits with 3 different values for the cooling time\,: from top to bottom
  $\tau_c=\{10, 2, 1\}/\Omega_p$. The shape of the cold region after encounter with the planet changes, as the coldest temperature corresponds to about 1 cooling time after the conjunction with the planet. { The boundary $T=T_{0}$ follows almost a straight line (white line on the plots) passing through the planet.  The line form an angle 
$\eta$ with the radial direction, scaling according to $\tan (\eta)\sim \tau_c$ (see text). } The resolution  is that of set  $B$(see text) for the top and  middle panels and that of $A$ for the bottom one.}
\label{fig:adc421}
\end{figure}

The high resolution allows to study the shape of this diagonal cold finger. In Fig.~\ref{fig:adc2}, the boundary $T=T_0$ follows almost a straight line passing through the planet, making an angle $\eta \sim 55^\circ$ with the radial direction. The relative velocity with respect to the planet of a particle orbiting around the star at a distance $r_p+\delta r$ being about 
$\delta r \times {3\over {2}}{ {\Omega_p}\over {r_p}}$, 
this line corresponds to a time $\delta t = \frac{2\,\tan\eta}{3\,\Omega_p} \propto \tau_c$ after the conjunction with the planet, that is after the maximal compression. In simulations where the cooling time is different, the angle $\eta$ changes accordingly, as can be seen in Fig.~\ref{fig:adc421}.

This shows that the particular structure we have discovered around low mass planets (this diagonal cold finger) is robust against the processes of heating and cooling. Provided the equation of state is not purely adiabatic, it will appear.

{  Since the gas cools  as it moves through the planet's potential the mechanism is optimally effective when the   cooling time is of the same order as the dynamical time. Actually, in Fig.\ref{fig:adc421}, top panel ($\tau_c = 10/\Omega_p$),  the diagonal cold finger is weaker and appears farther from the planet with respect to the cases with $\tau_c = {2,1}/\Omega_p$. We have 
   obtained a very weak cold finger for  a simulation run with    $\tau_c = 50/\Omega_p$ while the phenomenon didn't appear for a simulation run with  $\tau_c = 100/\Omega_p$.  \par The comparison with  the 3D radiative cooling rate
is not so straightforward. If we consider
 the radiative cooling time scale as defined  in \citet{BK10}:
\begin{equation}
\tau_{rad} = {H^2\over {D/c_v\rho}}
\label{taurad}
\end{equation}
were $D$ is the diffusion coefficient of Eq.\ref{energy3D} with $\lambda =1/3$.
The value of $\tau_{rad}\Omega_p$ is about 100 at $r=r_p$ and turns out to decrease for $r<r_p$ (see \citet{BK10}, Fig. A.1 for the same disc parameters as in our simulations). The phenomenon should therefore be even more effective going toward the inner part of the disc.}

\begin{figure}
\includegraphics[height=7truecm,width=7truecm]{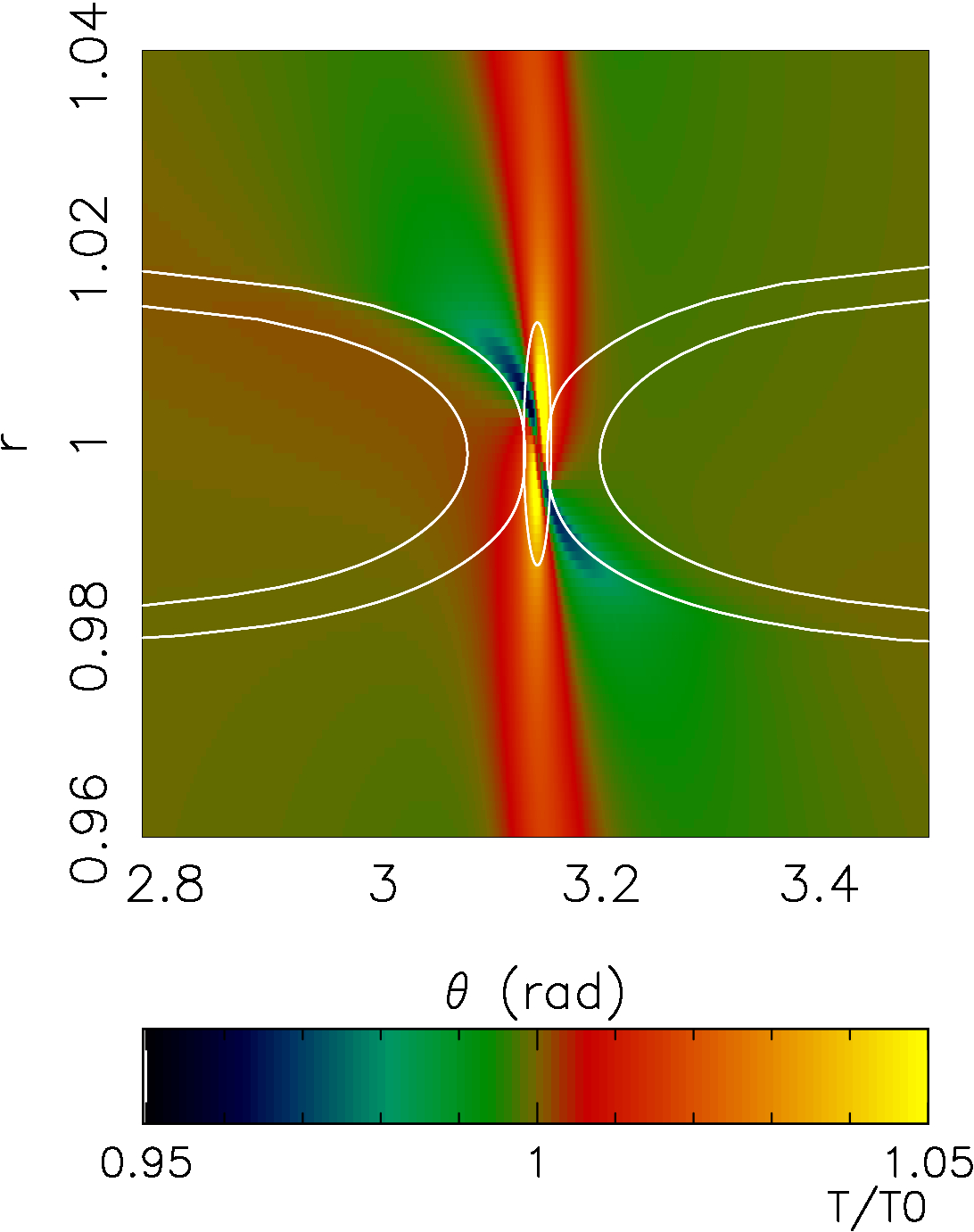}
\includegraphics[height=7truecm,width=7truecm]{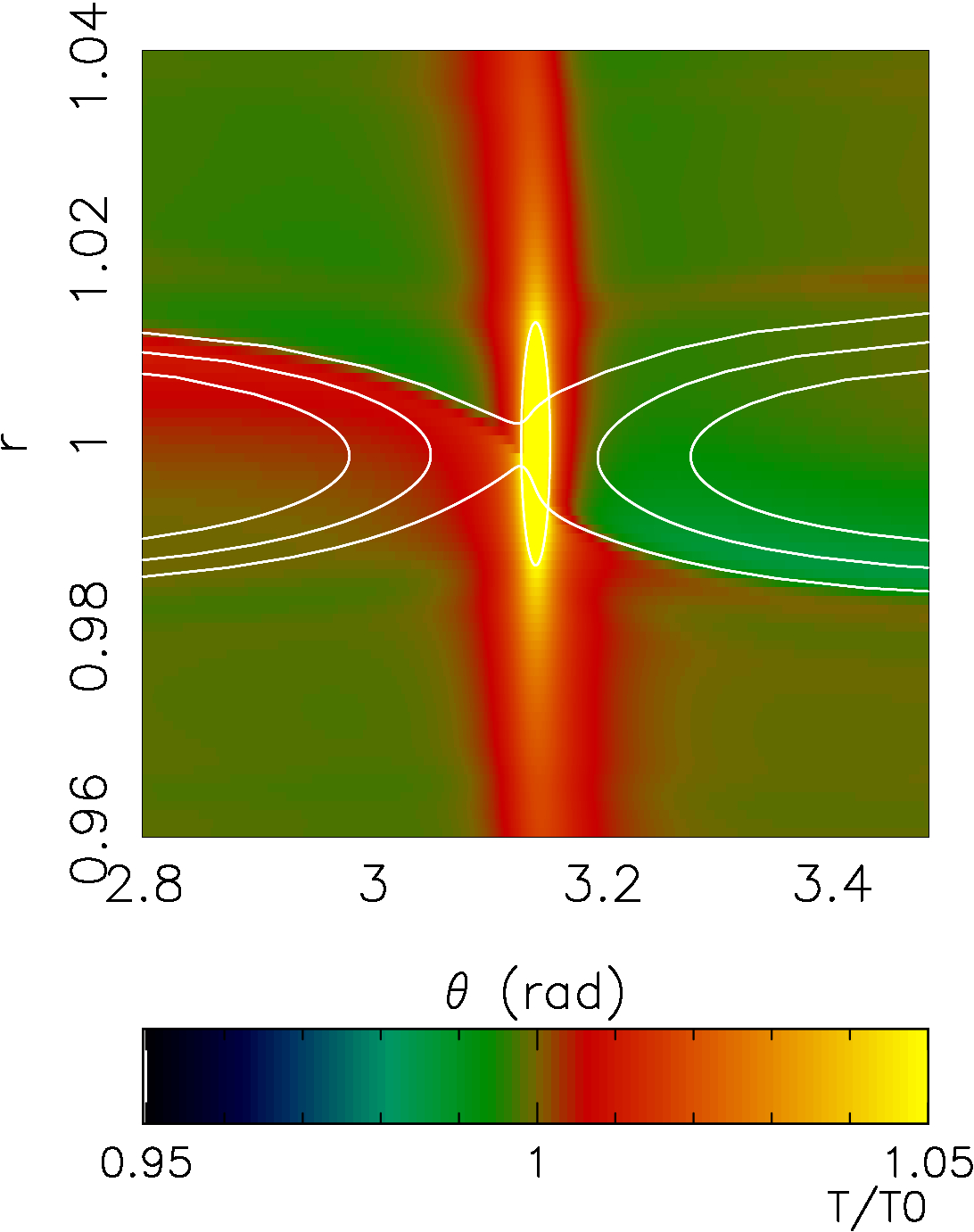}
\caption{{\bf 2D Simulation. Top panel}: perturbed temperature  around a $2 M_\oplus$ planet after 35 orbits, with $\tau_c=2/\Omega_p$. { Streamlines  making a U-turn far enough from the planet  to be heated by compression are associated to the usual thermal part of the corotation torque. Streamlines passing in the close vicinity of the planet show the cold finger resulting from  compressional heating and radiative cooling.} { {\bf Bottom panel}: perturbed temperature  around a $2 M_\oplus$ planet after 35 orbits in the adiabatic case.  The torque in the horseshoe region has not yet reached saturation so that  hot and cold plumes  are still present but no cold finger structure appears in this case.} }
\label{fig:adc4large}
\end{figure}

It should also be noted that the observed phenomenon is not in contradiction with the usual thermal part of the corotation torque \citep{PaardeMelle06,PaardeMelle08,BarMass08}. In Fig.~\ref{fig:adc4large}, top panel, the temperature variations are shown after 35 orbits, with $\tau_c=2/\Omega_p$, with a larger range in the azimuthal direction. The hot and cold plumes expected in the horseshoe region are present, along streamlines that did a U-turn far enough from the planet to avoid being heated significantly by compression (which would have made them lose internal heat). However, the cold finger is also present, and shows actually a larger temperature variation than the plumes. { For comparison we show (Fig.~\ref{fig:adc4large}, bottom panel), the same simulation with a pure adiabatic EOS (no cooling) after 35 orbits. The torque in the horseshoe region has not yet reached saturation so that  hot and cold plumes  are still present. The horseshoe region appears to be thinner than in the radiative case and circulation streamlines that pass through the high pressure region close to the planet   do not create  any cold finger. Actually, after the encounter with the planet the expansion cools the gas back at the unperturbed temperature.}  

\begin{figure}
\includegraphics[height=7truecm,width=7truecm]{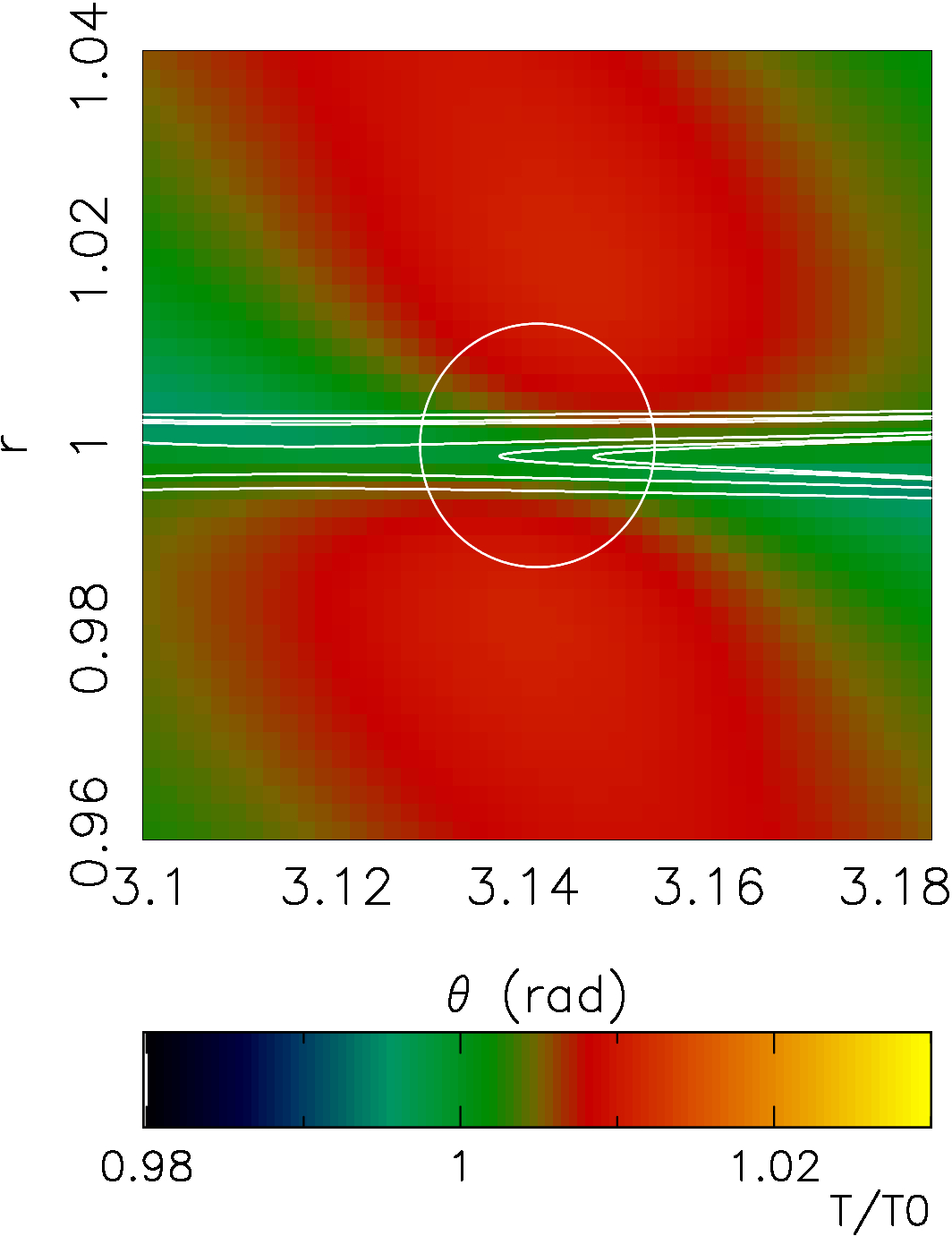}
\caption{{\bf 2D Simulation}: perturbed temperature  around a $2 M_\oplus$ planet after 60 orbits, with $\tau_c=4/\Omega_p$ and  standard $\epsilon$-potential with smoothing length  $\epsilon=0.6H$, where $H/r=0.05$.}
\label{fig:eps}
\end{figure}

{ In order to see the cold finger one needs  gas  passing through the high pressure region around the planet, without being bound to it. On the one hand
the high pressure region around the planet typically has the size of  its potential well: the Hill radius $R_H$ which is proportional to $m_p^{1/3}$. 
On the other hand an upper estimate of the region bound to the planet is 
the Bondi radius which is proportional to $m_p$. Therefore the cold finger can only appear if the Hill radius is significantly larger than the Bondi radius.
In our case,
\begin{equation}
{R_B \over R_H} = {{(m_p/7.1M_{\earth})^{2/3}} \over {(a/ 5.2AU)(T/ 75K)}}
\end{equation}
which implies that $R_B < R_H$ for $m_p < 7.1M_{\earth}$. 
Therefore, the cold finger effect was not prominent in previous 3D simulations 
because they have been done for planets more massive than $5M_{\earth}$ (\cite{PaardeMelle06,PaardeMelle08,KBK09}).
\par
Concerning 2D simulations,   this phenomenon was not observed before  because  
it is common to use a smoothing length of the order of the scale height of the disc (see Fig.\ref{fig:eps}). In fact, for a typical aspect ratio of  $5\%$, the  standard  smoothing length  $\epsilon=0.6H$ is larger than the Hill radius for planets smaller than $27M_{\earth}$.
Obviously, with a smoothing length comparable or larger than the Hill radius
it is not possible to simulate correctly the dynamics of the gas between the
Bondi radius and the Hill radius.
 In particular the pressure maximum is strongly weakened. We remind the reader that to see the cold finger effect in 2D simulations we used a smoothing length that is a fraction of the Hill radius. }

\section{On the asymmetry of the cold fingers}

We know from the measures of the radial torque density (Fig.\ref{normtorque})
that the cold finger appears to be asymmetric with a larger contribution at $r>1$ that is responsible for the negative total torque.

\begin{figure}
\includegraphics[height=6truecm,width=7truecm]{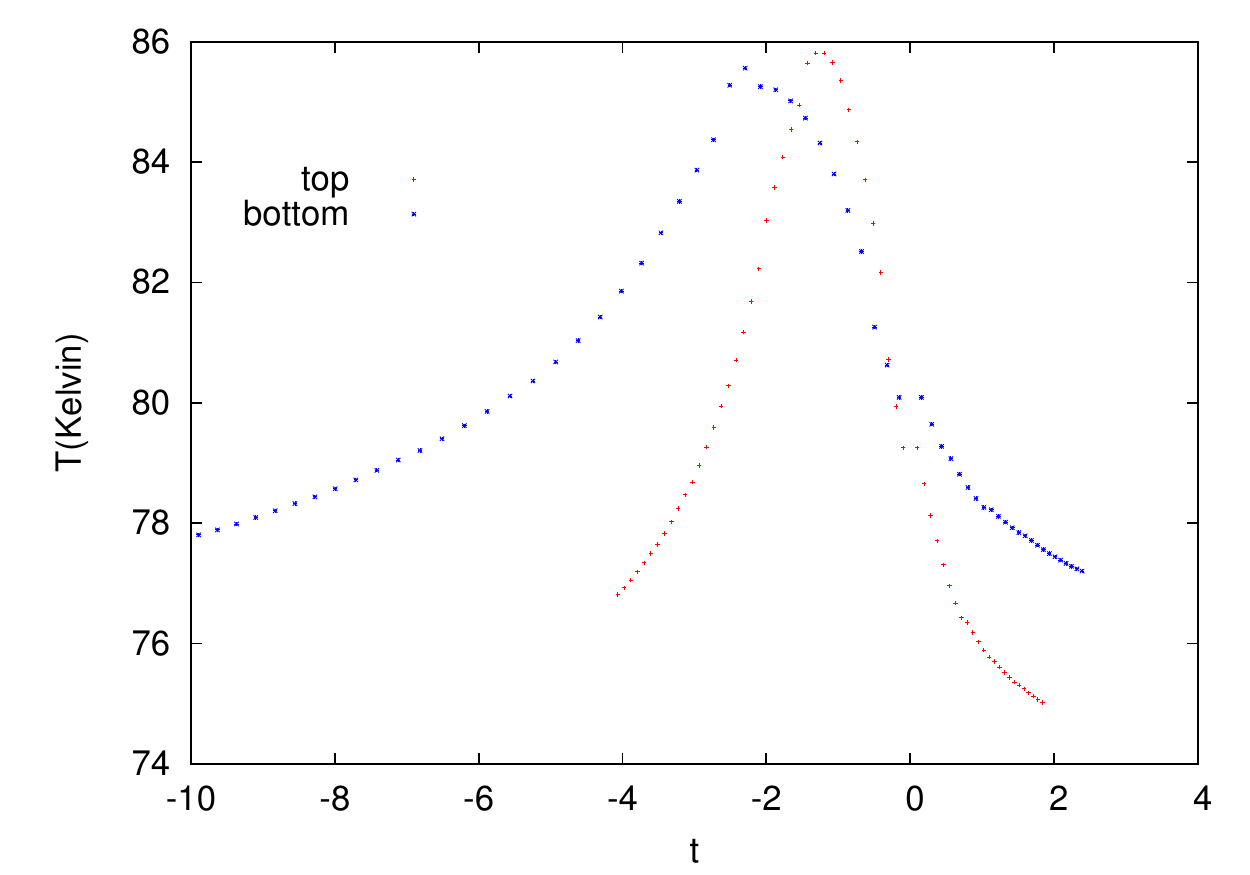}
\caption{Variation of the temperature along
 the two streamlines of
Fig.\ref{m2D}. The label 'top' corresponds to the streamline passing through $r-r_p= 5.5\,10^{-3}, \varphi-\varphi_p= -5.5\,10^{-3}$,  'bottom' to the one passing through   $r-r_p= -5.5\,10^{-3}, \varphi-\varphi_p=5.5\,10^{-3}$. 
Positive times correspond in both cases  to the part of the streamline
  downstream relative to the encounter of the gas with the planet. }
\label{streamm2T}
\end{figure}

In Fig.\ref{streamm2T} we show the  variation with time of the temperature  
associated to the 2 streamlines shown on Fig.\ref{m2D}, bottom panel. We recall that the gas moves from right to left for the top streamline and from left to right for the bottom one. In both cases the initial condition is chosen after the encounter with the planet: positive (negative) times in Fig.\ref{streamm2T} correspond to the part of the streamline
  downstream (upstream) relative to the encounter of the gas with the planet.
{ The gas associated to  the  top streamline (in Fig.\ref{m2D}, bottom panel)  has a larger relative temperature variation  with respect to the bottom one (in Fig.\ref{m2D}, bottom panel).Precisely, $75/76.8$ for top one instead of $77/77.75$ for the bottom one. Therefore the density enhancement for $r>1$ 
has to be sharper than for $r<1$.
} 

We consider the following explanation for the  asymmetry: 
  the flow is slightly  sub-Keplerian 
 due to the pressure gradient in the disc, therefore the stagnation points
are located at $r<r_p$ and as a consequence circulation streamlines pass closer to the planet location for $r>r_p$ than for $r<r_p$. 

\begin{figure}
\includegraphics[height=6truecm,width=7truecm]{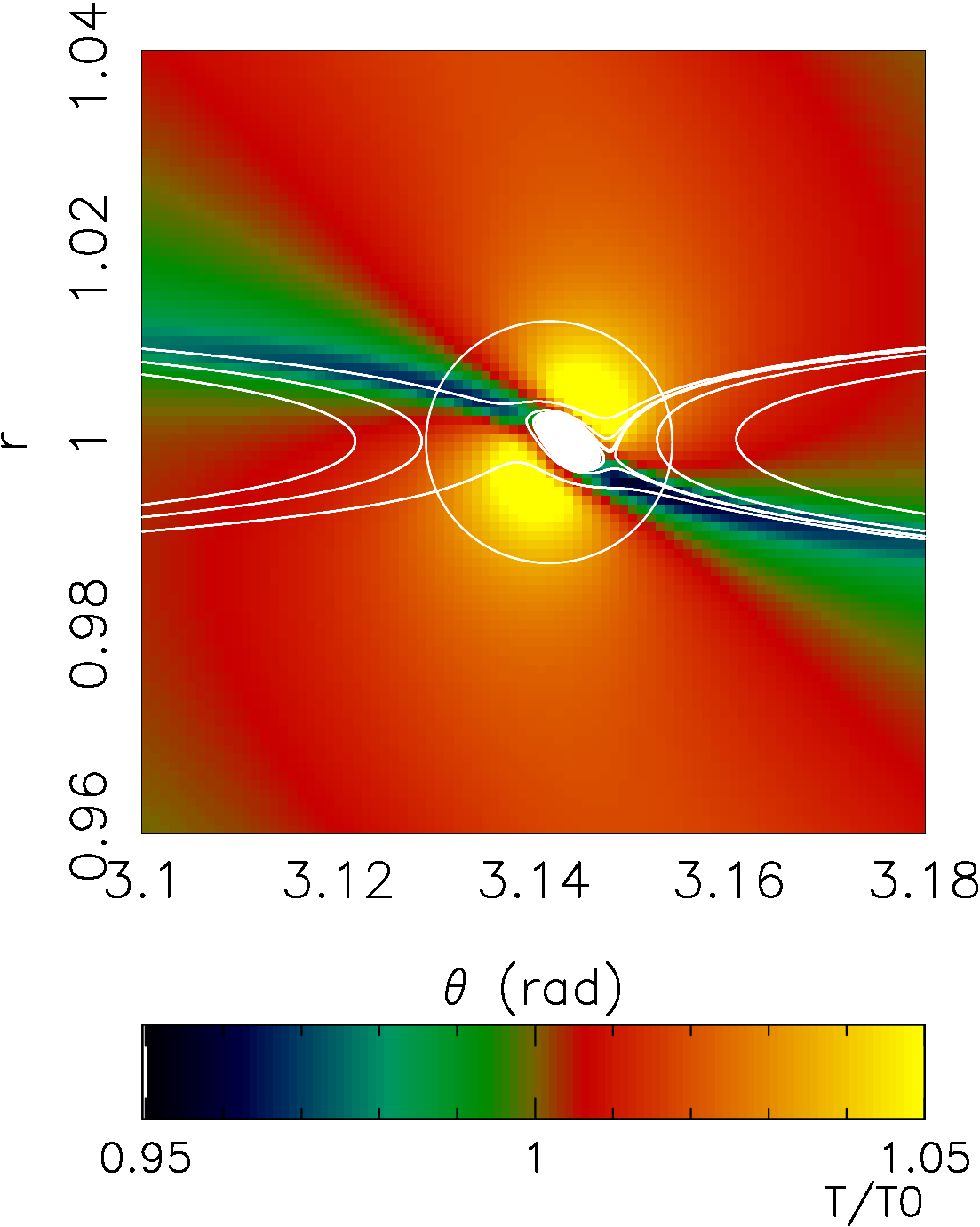}
\caption{{\bf 2D simulation}: perturbed temperature  around a $2 M_\oplus$ planet after 60 orbits for a simulation in which the gas has Keplerian speed, the surface density and the temperature are initially constant with respect to $r$.  }
\label{kepler}
\end{figure}

 In order to test this hypothesis, we have run a 2D simulation with  constant surface density  $\Sigma = \Sigma_0$ and disc height 
 $H(r)=h_0r^{3/2}$  in order to have constant temperature with respect to $r$,  so that the disc is Keplerian since the radial pressure gradient is null. Fig.\ref{kepler} shows the ratio $T/T_0$. When compared to Fig.\ref{fig:adc2}   the hot regions inside the Hill sphere (in yellow)
appear more symmetric with respect to the planet position but the cold finger is now slightly colder for $r<r_p$ than for $r>r_p$. 
We have to carefully investigate the topology of the streamlines around the planet to understand why the cold finger is not symmetric when the gas has Keplerian speed.
The streamlines appear to be sensitive on tiny variations in the velocity field 
and  not symmetric: in particular one streamlines  goes around the planet, and leave it on the bottom right. As a consequence, the cold finger is slightly colder on the bottom right than on the top left. 
{ The reason is that the viscous evolution of the disc makes the gas drift inwards.}
Notice that the variation of the angular momentum is proportional to the radial gradient of $\sqrt r \Sigma \nu$ so that, for constant viscosity, it is null for $\Sigma \propto 1/\sqrt r$,  as considered in the previous set of 2D simulations.  \par It is clear that a good knowledge of  the flows around the planet allows to determine the characteristics of the cold finger. Further studies are necessary to model the dynamics in the close vicinity of the planet.
This actually may well determine the migration direction and rate for small mass
planets.
\par

\section{Conclusion}
Using 3D hydrodynamical simulations including radiative transfer we have computed the torque acting on small
mass planets  kept on fixed orbits in the aim of finding  the critical mass for which the positive contribution
of the corotation torque becomes smaller than the Lindblad torque.
While we were expecting to observe  that the positive corotation torque fades away for decreasing mass of the planet,  we found a  new and unexpected feature:  a negative contribution to the torque that is not seen for planetary masses larger than $m_p \ga 5M_{\earth}$. 
 We explain this phenomenon observed in viscous discs \footnote{we do not refer to the nonviscous case nor to the limit of vanishing viscosity}  as follows. Small mass planets don't have a circum-planetary disc, but their gravitational force still increases the pressure and temperature in their neighborhood.  Looking at density and temperature fields it appears  that  some circulating or even some librating streamlines go through pressure increase as they encounter the planet. If the equation of state is locally isothermal or purely adiabatic, the temperature, density and pressure are identical before and after this pressure peak. However, if radiative transfer is included, the gas is cooler, thus denser, after the encounter with the planet. This leads to the formation of a cold and dense finger ahead of the planet just inside of its orbit, and behind the planet just outside of its orbit, outside of the horseshoe region. This phenomenon, that we have recovered with two dimensional high resolution simulations, explains the unexpected negative contribution
 observed on the torque just outside the planet orbit in 3D simulations.
\par
At our knowledge the cold finger was not observed in previous  numerical
computations. 
 We understand that, in order to have a torque contribution related to the cold finger in viscous discs, the following conditions must be satisfied:
\begin{itemize}
\item { planets have to be small enough  for their Hill radius to be significantly larger than their Bondi radius.}
\item the equation of state must be non  isothermal neither adiabatic, with an appropriate cooling time.
\item the planetary potential has to be close to the real one, for example the
  cubic potential introduced in \citet{KBK09} suited to 3D simulations.
\end{itemize}

The cold finger turns out not to be symmetric being more pronounced behind the planet with our standard disc's parameters, and it   appears to be responsible for the transition from outward to inward migration as the planet's mass decreases. However, this asymmetry is very sensitive to the properties of the flow around the planet. In particular, we found a positive total torque in a disc with flat surface density and temperature profile. Therefore, for specific disc properties, earth size planets could still migrate outwards, even when the usual
thermal contribution to the corotation torque fades. This enlightens the importance of the cold finger, and advocates for future studies of this new torque contribution.

\section{Appendix : Test calculations with FARGOCA}
The motion of the gas is described by the Navier-Stokes equations 
 in a rotating frame in spherical coordinates:
\begin{itemize}
\item {\bf Continuity equation}
\begin{equation}
\label{continuity}
{\partial \rho \over \partial t}+ \nabla \cdot (\rho \vec v)= 0
\end{equation}
where $\rho$ is the density of the gas and $v=(v_r,v_\varphi,v_\theta)$
the velocity, with $v_\varphi=r\sin (\theta)(\omega+\Omega)$ 
where $\omega$ is the azimuthal angular velocity in the rotating frame.
\item  {\bf Equations for the momenta.}
 The Navier-Stokes equations  for  the radial momentum  $J_r= \rho v_r$,
the polar momentum $J_\theta = \rho r v_\theta$ and the angular momentum
$J_\varphi = \rho r\sin(\theta) v_\varphi = \rho r^2 \sin^2\theta (\omega+\Omega)$ read:
\begin{equation}
\label{navstocons3D}
\left\lbrace \begin{array}{lll}
{\partial J_r \over \partial t}+ \nabla \cdot (J_r\vec v) & = & \rho [{v_\theta ^2 \over r }+ {v_\varphi^2\over r} -{\partial \Phi \over \partial r}
     \\
& & +{1\over \rho} (f_r-{\partial P \over \partial r})] \\
{\partial J_\theta \over \partial t}+ \nabla \cdot (J_\theta \vec v) & = &
 \rho r[{v_\varphi^2\cot(\theta)\over r} -{1\over r}{\partial \Phi \over {\partial \theta}}  \\ & & +{1\over \rho } 
 ( f_\theta -{1\over r }{\partial P \over {\partial \theta}} ) ] \\
{\partial J_\varphi \over \partial t}+ \nabla \cdot (J_\varphi \vec v) & = &
\rho r \sin (\theta) [-{1\over {r \sin \theta}}{\partial \Phi \over \partial \varphi}    \\ & & +{1\over \rho}(f_\varphi 
 -{1\over {r \sin \theta}}{\partial P \over \partial \varphi} )]
\end{array} \right.
\end{equation}

The function $f=(f_r,f_\varphi,f_\theta)$ is the divergence of the
stress tensor (see for example \citet{Tassoul78}).
The potential $\Phi$ acting on the disc consists of the contribution of
the star $\Phi_*=-GM_{*}/r$ and  planets $\Phi_p$, plus indirect terms that arise from the primary acceleration due to the planets’ and disc’s gravity. 
\end{itemize}
To test the implementation of FARGOCA we consider first the isothermal 
setup.  The computational domain $(r,\varphi,\theta)$  consists of an annulus of the protoplanetary disc extending from $r_{min}$ to $r_{max}$. In the vertical 
direction the annulus extends from $\theta_D$ to $\pi-\theta_D$, the midplane being at $\theta = \pi/2$.
We use a grid of $N_r\times N_\varphi\times N_\theta$ cells equally
  spaced in in $r,\varphi,\theta$.
\subsection{Setup}
We consider the gas at equilibrium in the gravitational field of the 
star, i.e. the motion is circular at constant height $z$.
The vertical structure of the disc is provided by
the  hydrostatic equilibrium in the thin disc approximation.
 The initial density has a  Gaussian profile: 
\begin{equation}
\rho (r,\theta) =  f(r) \exp(-{{r^2({\pi \over 2}-\theta)^2} \over {2H(r)^2}})
\end{equation}
where $H(r)$ is the height of the disc $H(r)=h_0r^{1+\beta}$
with $h_0$  the disc aspect ratio and  $\beta$ is the flaring index.
The radial coordinate in the code is normalized by $a_p$. 
The function $f(r)$ is found setting the vertical integration
of  $\rho$ equal to the disc surface density:
\begin{equation}
\Sigma(r)=\Sigma _0 r^{-\alpha_{\Sigma}}
\label{surfdens}
\end{equation}
 and reads:
\begin{equation}
f(r) = {\Sigma _0 \over h_0\sqrt {2\pi}} r^{-\alpha_{\Sigma}-1-\beta}
\end{equation}
Let's remark that in the thin disc approximation we use $\cos(\theta) \simeq ({\pi \over 2}-\theta)$ and $r\sin (\theta) \simeq r$.
The initial values of the velocities are $v_r=0$ and $v_\theta=0$,
 $v_\varphi$ is obtained from the balance of the radial forces.

The initial conditions are obtained through an approximation so that they do not
correspond exactly to the equilibrium, the gas is observed to relax toward a more stable state only marginally different from the initial configuration.\par
\subsection{Fargo Algorithm in the 3D case}
The specificity of the FARGO code  (Fast Advection in Rotating Gaseous Objects, \citet{Masset00}) is a much larger timestep than usual hydrodynamical codes. 
 The FARGO algorithm is particularly
 well-suited to the description of a Keplerian disc where
 the traditional Courant condition provides very small timesteps due to 
 the fast orbital motion at the inner boundary of the disc. In the FARGO algorithm, the timestep is limited by the perturbed velocity  arising from the differential rotation. 
In order to extend this procedure to the 3D case we consider  the
thin-disc approximation  and we compute the mean velocity:
\begin{equation}
\bar v_\varphi(i) = {1\over N_{\varphi}N_{\theta}}\sum _{j,h} v_\varphi(i,j,h)
\end{equation}
where the sums runs over the azimuthal and polar grid-cells.
The local variation from the mean flow, i.e. the  residual velocity,
 at each cell is then: 
$$ v^{res}_{i,j,h} =  v_\varphi(i,j,h)-\bar v_\varphi(i) $$
which provides at each cell a limitation timestep:
\begin{equation}
\delta t_{\varphi} = {r\sin(\theta)\Delta \varphi \over v^{res}_{i,j,h}}
\end{equation}
A normal CFL-condition (\citet{StoneNorman92}) taking into account $\delta t_{\varphi}$ provides the integration timestep.
\par
\subsection{Parallelization}
The use of a three dimensional code for the study of gas-disc interaction with small mass planets (less than 10 Earth masses) requires high resolution grids. 
 While in 2D the  parallel implementation of the code is in general not necessary it becomes very important in the 3D case. For the  specific nature of the FARGO algorithm the MPI is implemented by splitting the mesh  only radially in a number of rings equal to the number of MPI processes.  
The communications between neighboring processors is done only once per timestep in order to set the hydrodynamic variables in the ghost zones. The ghost zone
is made of $5\times N_s\times N_{\theta}$ grid-cells which is a relatively large number. Each MPI process has access to $N_r\times N_s\times N_{\theta}$ grid-cells with 
$N_r \geq 40$ being a good compromise between communication time and computation time.  The $N_r$ rings can then be further splitted on shared memory multi-core processors  using OpenMP instructions.
 We use in the following this hybrid parallelization which provided us an efficient code.

\subsection{Testing the isothermal configuration}
\begin{figure}

\includegraphics[height=3.5truecm,width=3.5truecm]{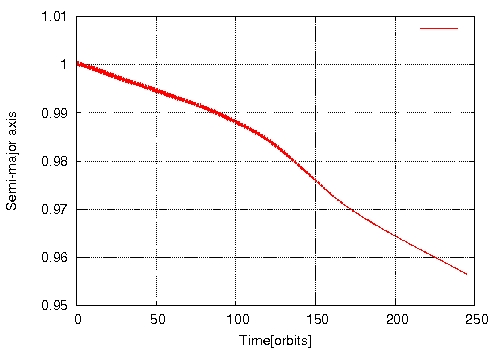}
\hskip -2 truecm
\includegraphics[height=3.5truecm,width=3.5truecm]{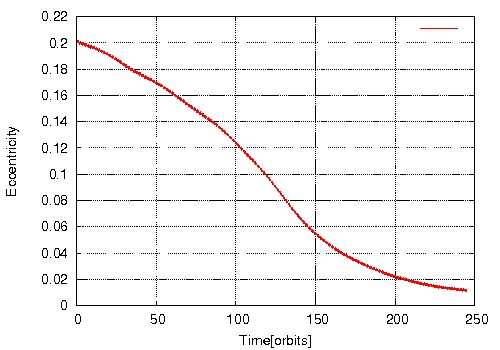}
\includegraphics[height=3.5truecm,width=3.5truecm]{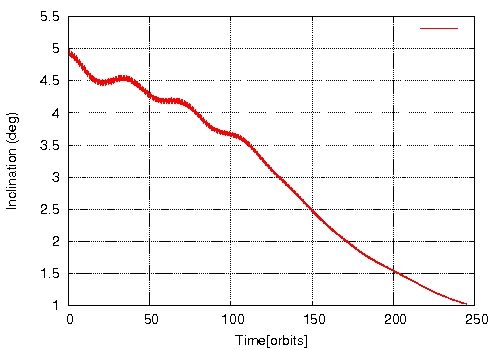}
\caption{Evolution of the semi-major axis, eccentricity and inclination with time for a planet with initial inclination: $i_0=5^\circ$ and initial eccentricity $e_0=0.2$.}
\label{ecceincli}
\end{figure}

In order to test the evolution of planet embedded a 3D disc, we have repeated some of the experiments in  \citet{Cresswell07}. The authors consider  the
evolution of a 20 Earth mass planet embedded in a 3D disc. 
They use for the gravitational potential of the planet the smoothed potential 
common in 2D simulations (Eq.\ref{smooth})
The planet is initially set at $a_p=5.2AU$ ($r=1$ in code units) in a disc of mass
$M= 0.07 M_*$ extending from
2.08AU to 13AU ($0.4<r<2.5$), with ${H\over r} = 0.05$.
The slope of the surface density is $\alpha_{\Sigma}=0.5$ and the flaring index
$\beta$ is set to zero. 
The value of $\theta_D $ is $75^\circ$ and  the smoothing length is $r_{\rm sm}=0.8R_H$. 
A Hill cut is applied when estimating the torque of the disc on the planet.
The resolution of the grid is $131\times 388 \times40$ ($N_r\times N_\varphi\times N_\theta$). For this kind of simulations we use rigid boundaries in the radial and vertical direction, the problem is periodic in azimuth.
In Fig.16 of \cite{Cresswell07} the evolution of an orbit with initial
inclination of $5^\circ$ and initial eccentricity of $0.2$ is studied.
Fig.\ref{ecceincli} shows the results obtained with FARGOCA which nicely agree with those of  \cite{Cresswell07}.
\subsection{Testing the fully radiative disc}
In order to test the fully radiative implementation we have repeated some
of the experiments presented in \cite{KBK09} with the same disc setup.
 In the full radiative set up the energy is
initialized as follows:
\begin{equation}
E(r,\theta) = {P(r,\theta)\over (\gamma-1)}={h_0^2\rho(r,\theta) \over (\gamma-1)} r^{-1-\alpha_{\Sigma}+2\beta} 
\end{equation}
Contrary to the isothermal case, we have no analytical prescription for an initial radiative equilibrium state and we first obtain it numerically.
The disc is  
axisymmetric  in absence of the planet, therefore we model
it in  2D, with  coordinates   $(r,\theta)$.

\begin{figure}
\includegraphics[height=6truecm,width=7truecm]{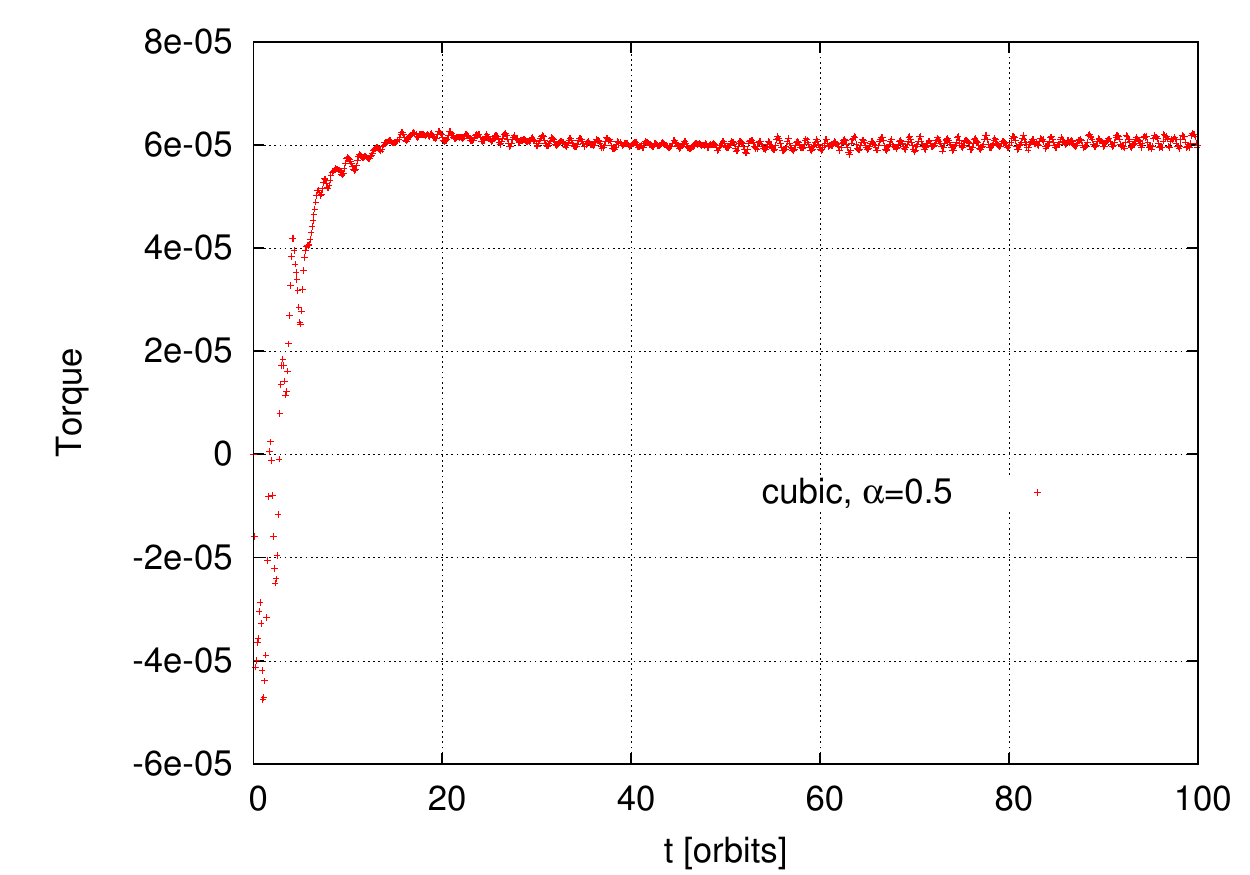}
\caption{Evolution of the total torque with time for a planet of $20 M_{\earth}$
 in the radiative case for the cubic potential of Eq.\ref{cubic} with $\alpha=0.5$.}
\label{ttm20}
\end{figure}

 For the set of parameters chosen for this simulation the balance
between viscous heating and radiative cooling reduces the aspect ratio of the disc from the initial value of $0.05$ to $0.037$ as explained in Section \ref{sect3}. Once the 2D equilibrium is achieved all the gas fields are expanded to 3D
and a planet of  $20 M_{\earth}$ is embedded  in such an equilibrium disc and
held on fixed  circular  orbits at $r=1$, $\varphi = \pi$, $\theta=\pi/2$ (midplane).
For this kind of simulations we use reflecting boundaries in the radial and vertical direction, the problem is periodic in azimuth.
 Fig.\ref{ttm20} shows the evolution with time of the total torque
acting on the planet (excluding of the inner part of the Hill sphere),
 for the cubic potential of Eq.\ref{cubic} with $\alpha=0.5$.
The torque becomes almost constant after about 40 orbits, with a positive value of $6\cdot 10^{-5}$ in good agreement with the results of \citet{KBK09} (their Fig.14, top). 

\section*{Acknowledgments}
We thanks F. Masset for useful discussions { and an anonymous 
referee for helping in  improving  the manuscript} 
The Nice group is thankful to ANR for supporting the MOJO project. The computations have been done on the ``Mesocentre SIGAMM" machine, hosted by the Observatoire de la C\^ote d'Azur.

\end{document}